\documentclass[a4paper,11pt]{article}
\pdfoutput=1 
\usepackage{jheppub} 
\usepackage[T1]{fontenc} 
\RequirePackage[displaymath]{lineno} 

\usepackage{epsfig}
\usepackage{graphicx}
\usepackage{dcolumn}
\usepackage{bm}
\usepackage{ltablex,booktabs}
\usepackage{overpic}
\usepackage{subfigure}
\usepackage{float}
\usepackage{color}
\usepackage{amsmath}
\usepackage{mathcomp}
\usepackage{mathrsfs}
\usepackage{multirow}
\usepackage{rotating}
\usepackage{amssymb}
\usepackage{gensymb}
\usepackage{amsmath}
\usepackage{tabularx}
\usepackage{epstopdf}

\title{Amplitude analysis and branching fraction measurement of the decay \boldmath $D_{s}^{+} \to K^+\pi^+\pi^-$}
\collaborationImg{\includegraphics[width=0.15\textwidth, angle=90]{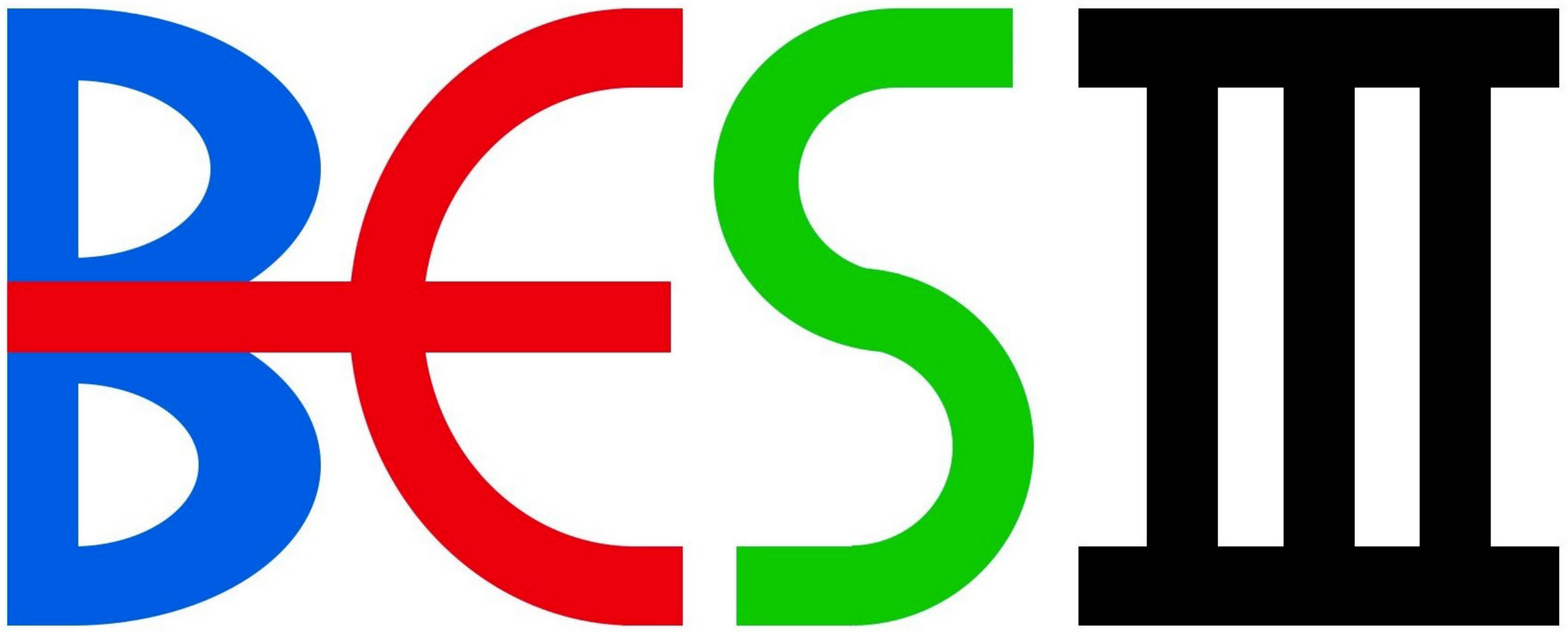}}
\collaboration{The BESIII Collaboration}

\date{\today}
\abstract{Using $6.32$ fb$^{-1}$ of $e^{+}e^{-}$ collision data collected at the center-of-mass energies between 4.178 and 4.226 GeV with the BESIII detector, 
we perform an amplitude analysis of the decay $D^+_s \to K^+\pi^+\pi^-$ and determine the amplitudes of the various intermediate states. The absolute branching fraction of $D^+_s\to K^+\pi^+\pi^-$ is measured to be ($6.11\pm0.18_{\rm stat.}\pm0.11_{\rm syst.})\times 10^{-3}$.
The branching fractions of the dominant intermediate processes $D_{s}^{+} \to K^+\rho^0, \rho^0 \to \pi^+\pi^-$ and $D_{s}^{+} \to K^*(892)^0\pi^+, K^*(892)^0 \to K^+\pi^-$ are determined to be $({{1.96}}\pm{{0.19}}_{\rm stat.}\pm{{0.23}}_{\rm syst.})\times 10^{-3}$ and $(1.85\pm{{0.12}}_{\rm stat.}\pm{{0.13}}_{\rm syst.})\times 10^{-3}$, respectively. The intermediate resonances $f_0(500)$, $f_0(980)$, and $f_0(1370)$ are
observed for the first time in this channel. }

\arxivnumber{}
\begin{document}
\maketitle
\flushbottom

\section{Introduction}
One popular approach for studies of hadronic charm decays involves application of approximate flavor symmetries, such as flavor ${\rm SU(3)}_F$~\cite{Ryd:2009uf}. However, the ${\rm SU(3)}_F$ flavor symmetry breaking effect has been observed in $D^0 \to K^+K^-$ and $D^0 \to \pi^+\pi^-$ for the first time, and later in the other singly Cabibbo-Suppressed (SCS) charm decays~\cite{PDG}. 
The SCS decay $D_s^+ \to K^+\pi^+\pi^-$, with low contamination from other charm decays, is a promising channel to study the ${\rm SU(3)}_F$ breaking effect. 
Furthermore, the measurements of the asymmetries of the branching fractions (BFs) of the charge conjugated decays of charmed mesons aid our understanding of charge-parity violation in the charm sector. To date, there have been a few measurements of charge-parity asymmetries, $A_{\mathit{{CP}}}$, in the SCS $D_s^\pm$ decay modes~\cite{CLEO:2013bae,PhysRevD.81.052013,2021}.

Two-body charmed meson decays $D_s^{\pm} \to VP$, where $V$ and $P$ denote vector and pseudoscalar mesons, respectively, have been studied in various approaches. The theoretical predictions of the BFs of the $D_s^+ \to K^+\rho^0$ ($\rho^0$ represents $\rho(770)^0$ throughout this paper) and $D_s^+ \to K^*(892)^0\pi^+$ processes are listed in Table~\ref{theoreticalresult}. References~\cite{Cheng:2016ejf,Wu:2004ht} 
studied these decay channels taking into account the ${\rm SU(3)}_F$ flavor symmetry breaking effect, while 
Ref.~\cite{PhysRevD.89.054006} {{uses}} a factorisation-assisted topological-amplitude approach with the $\rho$-$\omega$ mixing. 
Information about $D_s^+ \to K^0\rho^+$, $D_s^+ \to K^*(892)^0\pi^+$ and $D_s^+ \to K^*(892)^+\pi^0$ has been extracted from the decay $D_s^+ \to K_S^0\pi^+\pi^0$~\cite{2021}, but is inconclusive regarding these {{models}}. 
More measurements are needed to confront the theoretical predictions.

The CLEO collaboration has reported the absolute BF of  $D_s^+ \to K^+\pi^+\pi^-$ to be $(0.654\pm0.033_{ \rm stat.}\pm0.025_{\rm syst.})\%$~\cite{CLEO:2013bae}, using 600 $\rm{pb}^{-1}$ of $e^+e^-$ collisions recorded at a center-of-mass energy $\sqrt{s} = $ 4.17 GeV. An amplitude analysis of this channel has been performed by the FOCUS collaboration with $567\pm31$ signal events~\cite{FOCUS:2004muk}. Using 6.32$\rm\  fb^{-1}$ of $e^+e^-$ collision data collected with the BESIII detector at $\sqrt{s} = 4.178-4.226 {\rm \ GeV}$, we perform an amplitude analysis and BF measurement of the $D_s^+ \to K^+\pi^+\pi^-$ decay with the world's best precision. Charge conjugation is implied throughout this paper except when discussing $\mathit{{CP}}$ violation.

\begin{table}[htbp]\small
	\centering
	\begin{tabular}{l c c c c}
	\hline
		Channel
		&PDG~\cite{PDG}
		&\multicolumn{1}{c}{Cheng $et.\ al$~\cite{Cheng:2016ejf}}
		&\multicolumn{1}{c}{Wu $et.\  al$~\cite{Wu:2004ht}}
		&\multicolumn{1}{c}{Qin $et.\  al$~\cite{PhysRevD.89.054006}}\\
	\hline

	    $D_s^{+}\to K^+\rho^0$ &2.5 $\pm$ 0.4 &1.22 $\pm$ 0.06 &1.2  &2.5 \\
		$D_s^{+}\to K^*(892)^0\pi^+$ &2.13 $\pm$ 0.36 &2.06 $\pm$ 0.08 &3.3  &2.35 \\
        \hline
	\end{tabular}
	\label{theoreticalresult}
	\caption{The experimental measurements and theoretical predictions of the BFs of $D_s^+ \to K^+\rho^0$ and $D_s^+ \to K^*(892)^0\pi^+$ ($\times 10^{-3}$).}
\end{table}

\section{Detector and data sets}
\label{sec:detector_dataset}

The BESIII detector is a magnetic spectrometer~\cite{ABLIKIM2010345,Ablikim_2020} located at the Beijing Electron Positron Collider (BEPCII)~\cite{Yu:IPAC2016-TUYA01}. A helium-based multilayer drift chamber (MDC), a plastic scintillator time-of-flight system (TOF), and a CsI(Tl) electromagnetic calorimeter (EMC) compose the cylindrical core of the BESIII detector, and they are all enclosed in a superconducting solenoidal magnet providing a 1.0 T magnetic field. The solenoid is supported by an octagonal flux-return yoke with resistive plate counter muon identifier modules interleaved with steel. The acceptance of charged particles and photons is 93\% over a 4$\pi$ solid angle. The charged-particle momenta resolution at 1.0 GeV/$c$ is 0.5\%, and the specific energy loss ($dE/dx$) resolution is 6\% for the electrons from Bhabha scattering. The EMC measures photon energies with a resolution of 2.5\%(5\%) at 1 GeV in the barrel (end-cap) region. The time resolution of the TOF barrel part is 68 ps, while that of the end-cap part is 110 ps. The end-cap TOF was upgraded in 2015 with multi-gap resistive plate chamber technology, providing a time resolution of 60 ps~\cite{etof1,etof2,etof3}.  About 83\% of the data used in this paper benefits from this upgrade.  

Data samples corresponding to a total integrated luminosity of 6.32 f$\rm b^{-1}$ are used in this analysis. The integrated luminosities of the data samples taken at different energy points are listed in Table~\ref{energe}~\cite{XYZLumi,lumin,centermass}. These samples are classified into three sample groups, 4.178, 4.189-4.219, and 4.226 GeV according to the years of data taking and their running conditions. Since the $D_s^{*\pm}$ decays to $\gamma D_s^{\pm}$ and $\pi^0 D_s^\pm$ with BFs of (93.5$\pm$0.7)\% and (5.8$\pm$0.7)\%~\cite{PDG}, respectively, the signal events discussed in this paper are selected from the process $e^+e^-\to D_s^{*\pm}D_s^{\mp}\to\gamma D_s^+D_s^-$.

 \begin{table}[htb]
 \renewcommand\arraystretch{1.25}
 \centering

 \begin{tabular}{ccc}
 \hline
 $\sqrt{s}$ (GeV) & $\mathcal{L}_{\rm int}$ (pb$^{-1}$) & $M_{\rm rec}$ (GeV/$c^2$)\\
 \hline
  4.178 & 3189.0$\pm$0.2$\pm$31.9 & [2.050, 2.180] \\
  4.189 &  526.7$\pm$0.1$\pm$ 2.2 & [2 048, 2.190] \\
  4.199 &  526.0$\pm$0.1$\pm$ 2.1 & [2.046, 2.200] \\
  4.209 &  517.1$\pm$0.1$\pm$ 1.8 & [2.044, 2.210] \\
  4.219 &  514.6$\pm$0.1$\pm$ 1.8 & [2.042, 2.220] \\
  4.226 & 1056.4$\pm$0.1$\pm$ 7.0 & [2.040, 2.220] \\
  \hline
 \end{tabular}
 \label{energe}
  \caption{The integrated luminosities ($\mathcal{L}_{\rm int}$) and the 
   requirements on $M_{\rm rec}$ for various center-of-mass  energies. 
   The first and second uncertainties are statistical and systematic, 
   respectively. The definition of $M_{\rm rec}$ is given 
   in Eq.~\ref{eq:mrec}. }
\end{table}
Simulated inclusive Monte Carlo (MC) samples, forty times larger than the data sets, are produced  with a {\sc geant4}-based~\cite{GEANT4:2002zbu} MC simulation package, 
which includes the geometric description of the BESIII detector and the detector response, and are used to determine detection efficiencies and to estimate backgrounds.
The production of open charm processes, the initial-state radiation production of vector charmonium(-like) states and the continuum processes incorporated in {\sc kkmc}~\cite{Jadach:2000ir, Jadach:1999vf} are included into the samples. 
The known decay modes are modeled with {\sc evtgen}~\cite{Lange:2001uf, EVTGEN2} using BFs taken from the Particle Data Group (PDG)~\cite{PDG}, and the remaining unknown charmonium decays are modeled with {\sc lundcharm}~\cite{Chen:2000tv, LUNDCHARM2}. 
Final state radiation from charged final state particles is incorporated using {\sc photos}~\cite{PHOTOS}. 

\section{Event selection}
\label{ST-selection}

The tag method~\cite{MARK-III:1985hbd} is employed to select clean signal samples of $e^+e^- \to D_s^{*\pm}D_s^\mp \to \gamma D_s^+D_s^-$ in the following analyses. In this method,
a single-tag (ST) candidate requires a reconstructed $D_{s}^{-}$ decay to any of the ten hadronic final states listed in Table~\ref{tab:tag-cut}. A double-tag (DT) candidate requires that the $D^+_s$ is reconstructed in the signal mode $D_{s}^{+} \to K^+\pi^{+}\pi^{-}$ in addition to the $D_{s}^{-}$ decay to one of the tag modes. The selection criteria described here are the {{common}} requirements for both amplitude analysis and {{BF}} measurement. Further requirements for amplitude analysis and {{BF}} measurement are discussed in Sec.~\ref{AASelection} and Sec.~\ref{BFSelection}, respectively.

\begin{table}[http]
 \renewcommand\arraystretch{1.25}
 \centering

	 \label{tagwindow}
     \begin{tabular}{lc}
        \hline
        Tag mode                                     & Mass window (GeV/$c^{2}$) \\
        \hline
        $D_{s}^{-} \to K_{S}^{0}K^{-}$               & [1.948, 1.991]            \\
        $D_{s}^{-} \to K^{-}K^{+}\pi^{-}$            & [1.950, 1.986]            \\
        $D_{s}^{-} \to K_{S}^{0}K^{-}\pi^{0}$        & [1.946, 1.987]            \\
        $D_{s}^{-} \to K^{+}K^{-}\pi^{-}\pi^{0}$     & [1.947, 1.982]            \\
		$D_{s}^{-} \to K_{S}^{0}K^{-}\pi^{-}\pi^{+}$ & [1.958, 1.980]            \\
        $D_{s}^{-} \to K_{S}^{0}K^{+}\pi^{-}\pi^{-}$ & [1.953, 1.983]            \\
        $D_{s}^{-} \to \pi^{-}\pi^{-}\pi^{+}$        & [1.950, 1.987]            \\
        $D_{s}^{-} \to \pi^{-}\eta$   & [1.930, 2.000]            \\
        $D_{s}^{-} \to \pi^{-}\pi^{0}\eta$   & [1.920, 2.000]            \\
        $D_{s}^{-} \to \pi^{-}\eta^{\prime}$ 
				                                             & [1.938, 1.997]            \\
		
        \hline
      \end{tabular}
 \caption{Requirements on $M_{\rm tag}$ for {{the ten}} tag modes.}\label{tab:tag-cut}
\end{table}

All charged tracks reconstructed in the MDC must satisfy $|$cos$\theta|<0.93$, where $\theta$ is the polar angle with respect to the direction of the positron beam.
For charged tracks not originating from $K_S^0$ decays, the closest distance to the interaction point is required to be less than $\pm$10~cm along the beam direction and less than 1~cm in the plane perpendicular to the beam.
Particle identification (PID) for charged tracks combines the measurements of the $dE/dx$ in the MDC and the flight time in the TOF to form probabilities $\mathcal{L}(h)~(h=K,\pi)$ for each hadron ($h$) hypothesis.
The charged tracks are assigned as kaons or pions if their probabilities satisfy one of the two hypotheses, $\mathcal{L}(K)>\mathcal{L}(\pi)$ or $\mathcal{L}(\pi)>\mathcal{L}(K)$, respectively. 

The $K_{S}^0$ candidates are selected from all pairs of tracks with opposite charges whose distances to the interaction point along the beam direction are less than 20 cm. The selected tracks are assigned as $\pi^\pm$ and no further PID requirements are applied.
A primary vertex and a secondary vertex are reconstructed, and the decay length between the two vertexes is required to be greater than twice its uncertainty. 
The $K_{S}^0$ candidates {{are required to have}} a $\pi^{+}\pi^{-}$ invariant mass ($M_{\pi^{+}\pi^{-}}$) in the range $[0.487, 0.511]$ GeV$/c^{2}$.

The photon candidates are selected using the EMC showers. The minimum deposited energy of each shower in the barrel region~($|\!\cos \theta|< 0.80$) and in the end-cap region~($0.86 <|\!\cos \theta|< 0.92$) must be greater than 25 MeV and 50 MeV, respectively.  
The opening angle between the location of each shower in the EMC and the extrapolated position of the closest charged track must be greater than 10 degrees to reject showers originating from charged tracks. 
The shower is required to start within [0, 700]\,ns from the event time to suppress electronic noise and showers unrelated to the event.

The $\pi^0$ and $\eta$ candidates are reconstructed from  photon pairs with invariant masses in the ranges $[0.115, 0.150]$~GeV/$c^{2}$ and $[0.500, 0.570]$~GeV/$c^{2}$, respectively, which correspond to about three standard deviations of the invariant mass resolutions. To improve their invariant mass resolutions, we require that at least one photon comes from the barrel region of the EMC. A kinematic fit constraining the $\gamma\gamma$ invariant mass to the $\pi^{0}$ or $\eta$ known mass~\cite{PDG} is performed. The $\chi^2$ of the kinematic fit is required to be less than 30. 
The $\eta^{\prime}$ candidates are formed from $\pi^{+}\pi^{-}\eta$ combinations with an invariant mass within the range of $[0.946, 0.970]$~GeV/$c^{2}$.

Ten tag modes are reconstructed and the corresponding mass windows on the tag $D_{s}^{-}$ mass~($M_{\rm tag}$) are listed in Table~\ref{tab:tag-cut}~\cite{2021}. 
The quantity $M_{\rm rec}$ is defined as
\begin{eqnarray}
\begin{aligned}
	\begin{array}{lr}
	M_{\rm rec} = \sqrt{\left(E_{\rm cm} - \sqrt{|\vec{p}_{D_{s}^-}|^{2}+m_{D_{s}^-}^{2}}\right)^{2} - |\vec{p}_{D_{s}^-} | ^{2}} \; , \label{eq:mrec}
		\end{array}
\end{aligned}\end{eqnarray}
where $E_{\rm cm}$ is the energy of the initial state measured from the beam energy, $\vec{p}_{D_{s}^-}$ is the three-momentum of the $D_{s}^{-}$ candidate in the $e^+e^-$ center-of-mass frame and $m_{D_{s}^-}$ is the nominal $D_{s}^{-}$ mass~\cite{PDG}.
The mass windows of $M_{\rm rec}$ for $D_{s}^{-}$ candidates at each center-of-mass energy are listed in Table~\ref{energe} and help to suppress backgrounds that come from non-$ D_s^{*\pm}D_s^{\mp}$ processes~\cite{BESIII:2021anh}.

\section{Amplitude analysis}
\label{Amplitude-Analysis}
\subsection{Event selection}
\label{AASelection}
Further selection criteria used to improve the signal purity 
for the amplitude analysis are described next.  
These criteria will not be used in the BF measurement.

An extra photon candidate for the process $D_s^{*\pm} \to \gamma D_s^\pm$ 
(satisfying the requirements in Sec.~\ref{ST-selection}) is selected in order 
to perform a six-constraint (6C) kinematic fit.
This fit constrains the $D_s^-$ decay to the utilized tag mode and the 
$D_s^+$ decay to the signal mode with two hypotheses: the signal $D_s^+$ 
comes from a $D^{*+}_s$ or the tag $D_s^-$ comes from a $D^{*-}_s$. 
The total four-momentum is constrained to the initial four-momentum of the $e^+e^-$ system and the invariant masses of tag $D^-_s$ and $D^{*\pm}_s$ candidates are constrained to the corresponding nominal masses. 
The combination with the minimum $\chi^2$ is chosen. A $\chi^2_{\rm 6C} < 200$ requirement is applied to suppress background from other processes.
To ensure that all events fall into the phase-space (PHSP) boundary, 
a mass constraint on the signal $D^+_s$ is added to the 6C kinematic fit. 
The four-momenta of the final state particles 
from this new 7C kinematic fit are used in the amplitude analysis.

The $D_s^+ \to K^0_S(\to \pi^+\pi^-)K^+$ candidates are the dominant background for the $D_s^+ \to K^+\pi^+\pi^-$ decay. These backgrounds are rejected by requiring $M_{\pi^+\pi^-}$ being outside the mass interval of [0.4676,0.5276] GeV$/c^2$.

\begin{figure*}[!htbp]
  \centering
    \includegraphics[width=7.cm]{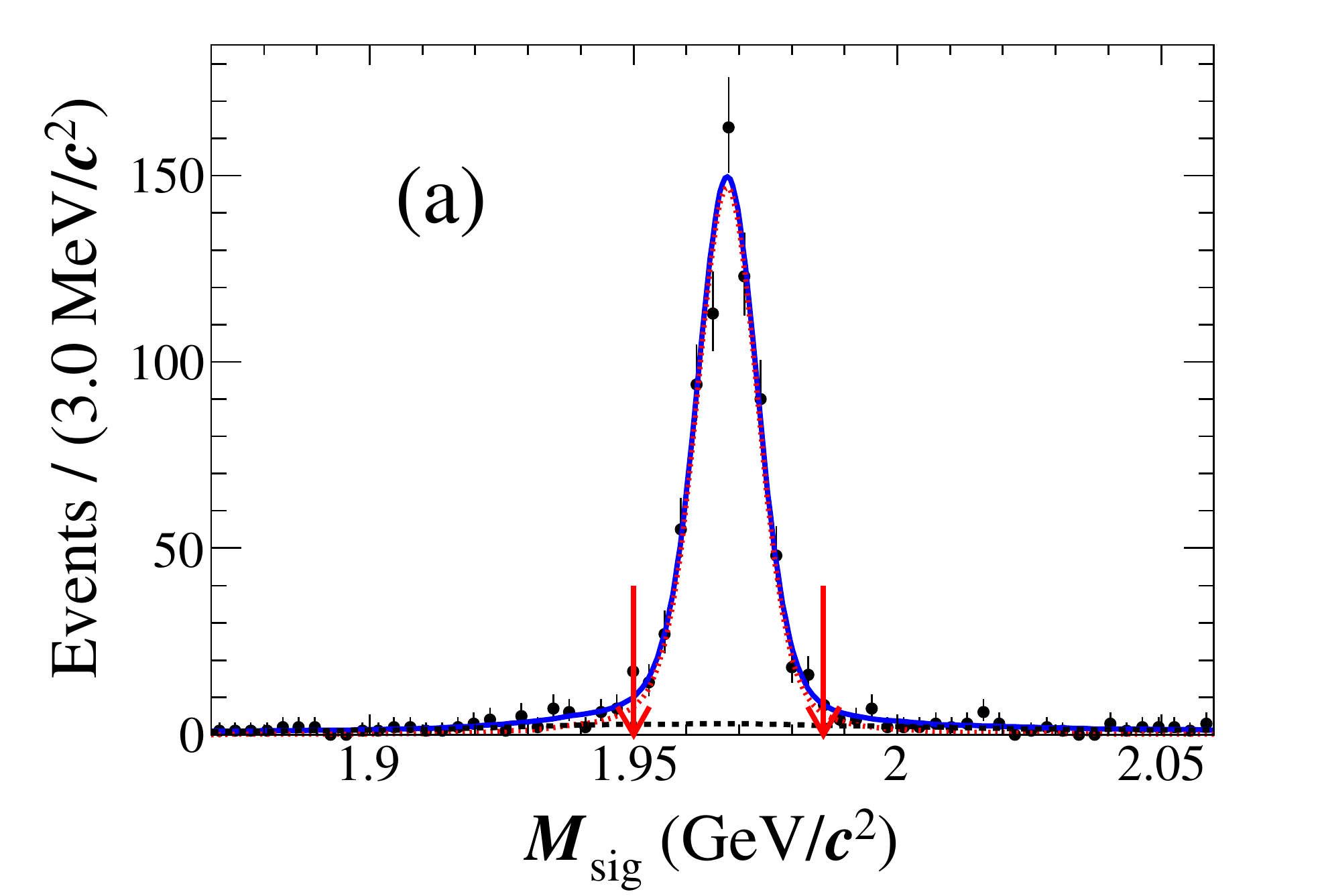}
    \includegraphics[width=7.cm]{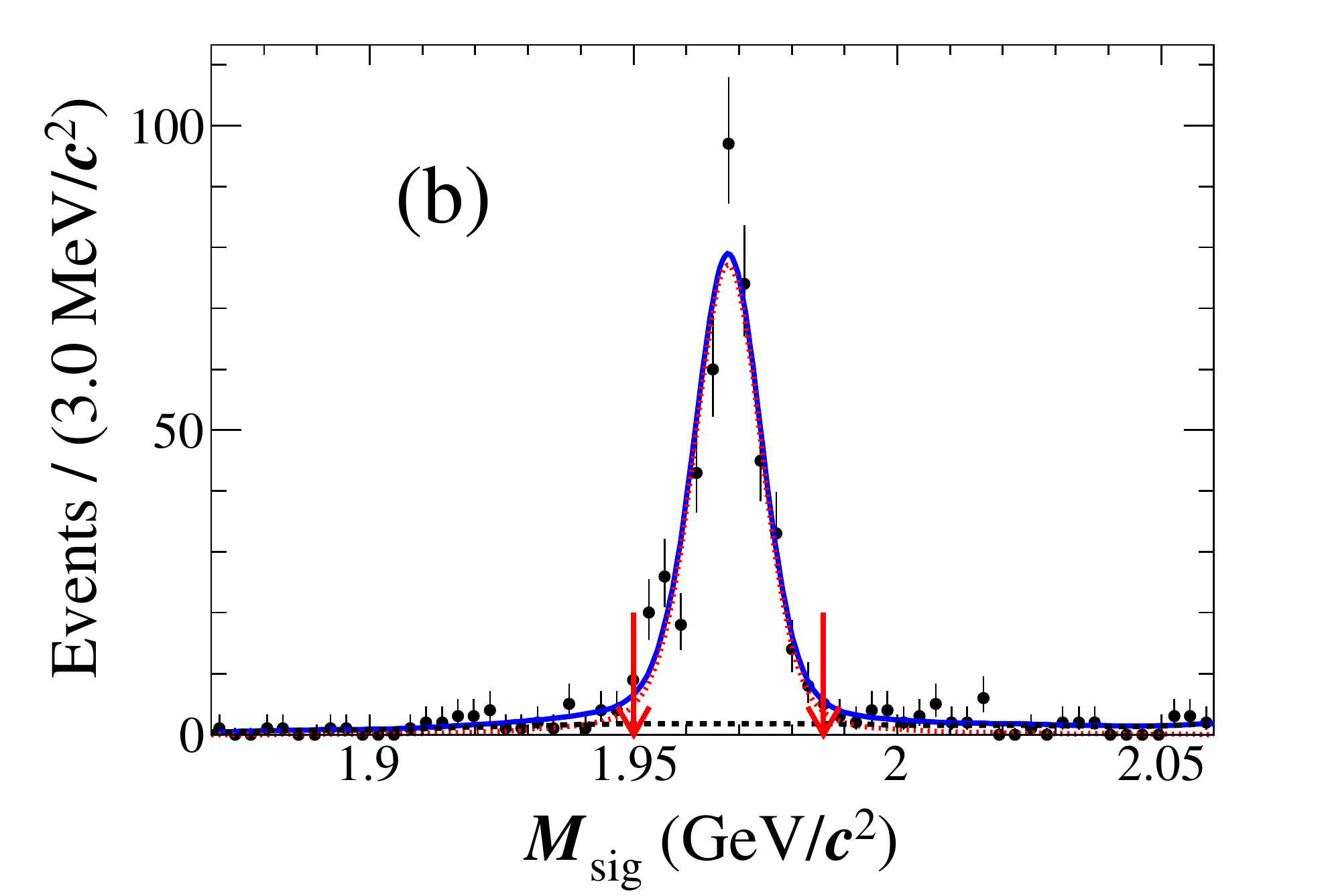}
    \includegraphics[width=7.cm]{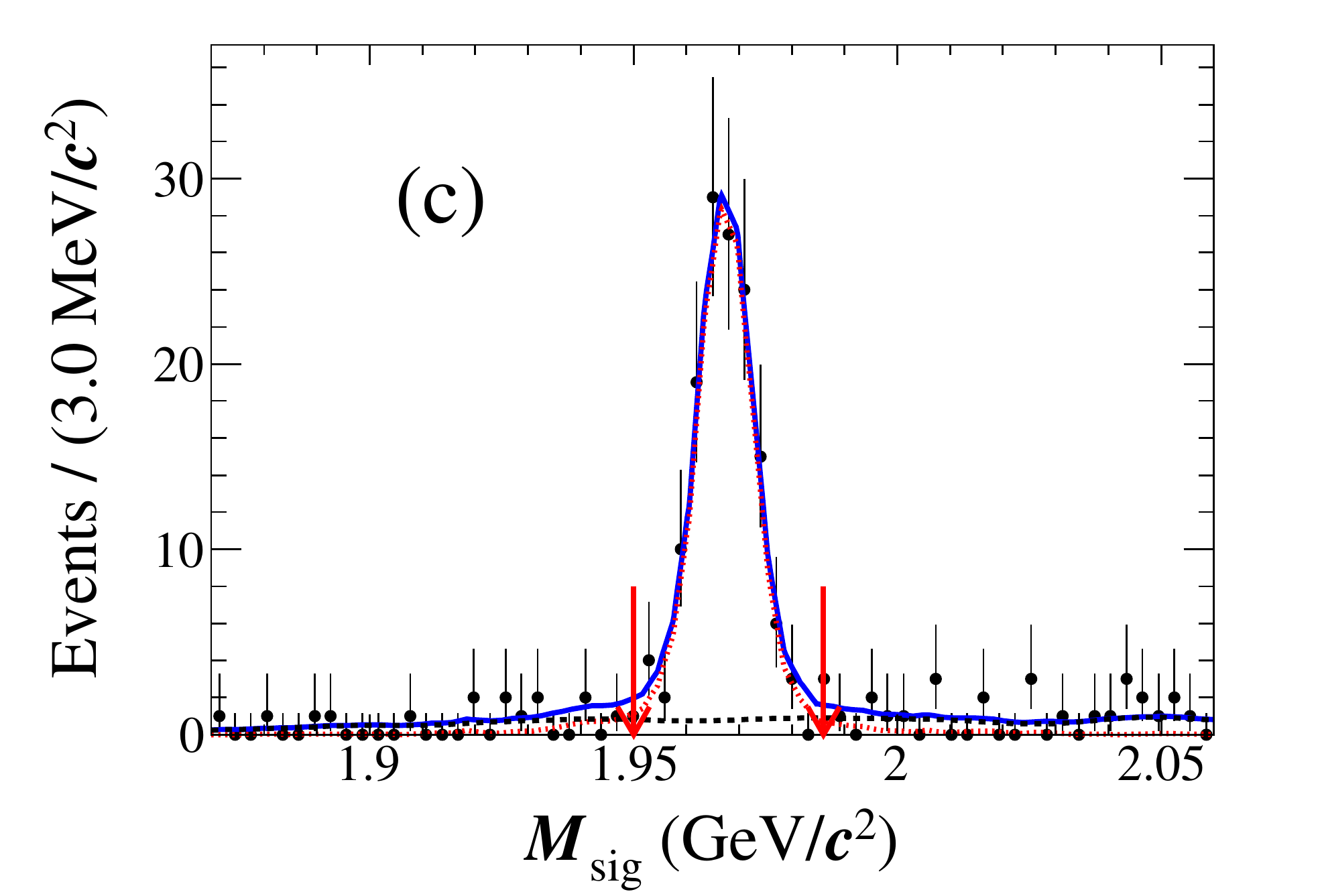}
    \includegraphics[width=7.cm]{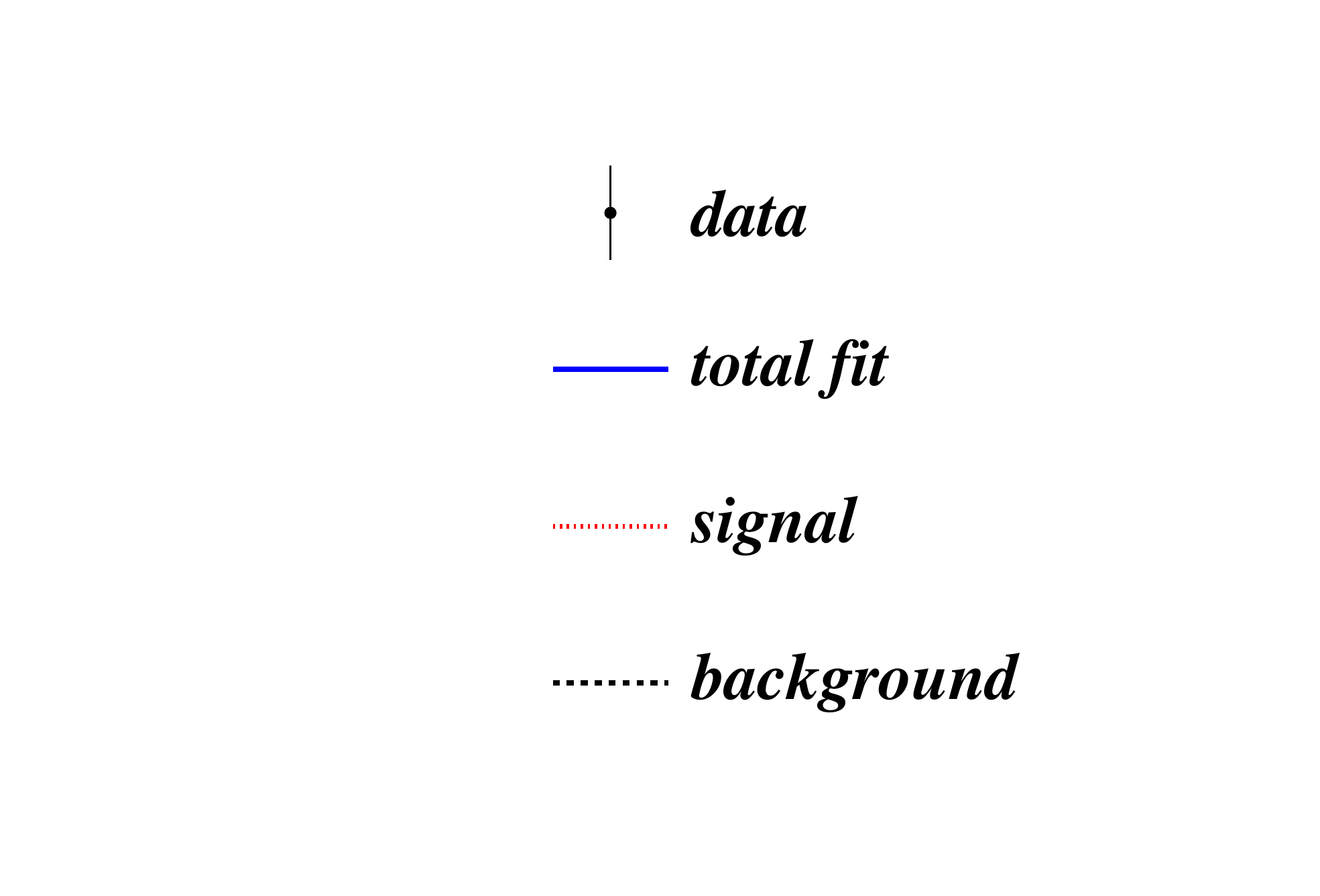}
  \caption{
    Fits to the $M_{\rm sig}$ distributions of the accepted signal candidates from the data samples at $\sqrt{s}=$ (a) 4.178~GeV, (b) 4.189-4.219~GeV and (c) 4.226~GeV. 
		The black points with error bars are data. 
		The blue solid lines are the total fits. 
		The red dotted and black dashed lines are the fitted signals and backgrounds, respectively.    The pairs of red arrows indicate the signal regions.
  } \label{fig:fit_Ds}
\end{figure*}

The fits to the invariant-mass distributions of the selected signal $D_s^{+}$ candidates ($M_{\rm sig}$) for various data samples are shown in Fig.~\ref{fig:fit_Ds}.
In the fits, the signal is described by the MC-simulated shape convolved with a Gaussian function describing the data-MC resolution difference, and the background is described with the shape derived from the inclusive MC sample.
Requiring $M_{\rm sig}$ $\in$ $[1.950, 1.986]$~GeV/$c^2$, 
we obtain 772, 444 and 140 signal candidates with purities of $(95.7\pm0.7)\%$, $(95.2\pm1.0)\%$, and $(92.6\pm2.2)\%$ for the data samples at $\sqrt{s}=4.178$, $4.189$-$4.219$, and $4.226$~GeV, respectively.

\subsection{Fit method}
The amplitude analysis of $D_s^{+}\to K^+\pi^+\pi^-$ is performed by an unbinned maximum likelihood fit. The likelihood function $\mathcal{L}$ is constructed with a signal-background combined probability density function (PDF). The log-likelihood is written as
\begin{equation}
  \ln{\mathcal{L}} = \begin{matrix}\sum\limits_{i=1}^{3} \sum\limits_{k}^{N_{D,i}} \ln [\omega^if_{S}(p^{k})\end{matrix}+(1-\omega^i)f_B(p^k)],  \label{loglikelihood}
\end{equation}
where $i$ indicates the data sample groups. The $p^k$ denote the four-momenta of the final state particles $K^+$, $\pi^+$, and $\pi^-$, where $k$ denotes the $k^{\rm th}$ event in data $i$.
The $N_{D,i}$ are the number of candidates in data $i$, $f_S(f_B)$ is the signal (background) PDF and the $\omega^i$ are the purities of the signals discussed in Sec.~\ref{AASelection}.

The signal PDF $\it{f_S(p)}$ is given by 
\begin{equation}
\begin{aligned}
    \it{f_S}(p) = \frac{\epsilon(p)|\mathcal{M}(p)|^{\rm{2}}R_{\rm{3}}(p)}{\int \epsilon(p)|\mathcal{M}(p)|^{\rm{2}}R_{\rm{3}}(p)dp},
\label{pwa:pdf}
\end{aligned}
\end{equation}
where $\epsilon(p)$ is the detection efficiency and $R_3(p)$ is the three-body PHSP function. The $R_3(p)$ is defined as
\begin{equation}
R_3(p) = \delta\left(p_{D_s}-\sum\limits_{j=1}^{3}p_j\right)\prod\limits_{j=1}^{3}\delta\left(p_j^2-m^2\right) \; \theta(E_j),
\label{pwa:R3}
\end{equation}
where $j$ runs over the three daughter particles, $E_j$ is the energy of particle $j$ and $\theta(E_j)$ is the step function. The total amplitude $\mathcal{M}$ is treated with the isobar model,
which uses the coherent sum of the amplitudes of the intermediate processes, $\mathcal M(p) = \sum{c_n\mathcal A_n(p)}$,
where $c_n = \rho_ne^{i\phi_n}$ is the corresponding complex coefficient. The magnitude $\rho_n$ and phase $\phi_n$ are the free parameters in the fit.
The amplitude of the $n^{\rm th}$ intermediate state ($\mathcal A_n$) is
\begin{equation}
	\mathcal A_n(p) = P_n(p)S_n(p)F^r_n(p)F^D_n(p).
\end{equation}
Here, $P_n(p)$ is the propagator of the intermediate resonance, $S_n(p)$ is the spin factor, $F^{r}_n(p)$ and $F^{D}_n(p)$ are the Blatt-Weisskopf barrier factors for the intermediate resonance and $D_s^+$, respectively.

The background PDF is given by
\begin{equation}
    f_B(p)=\frac{B(p)R_3(p)}{\int{\epsilon(p)B_{\epsilon}(p)R_3(p)}dp},	
  \label{bkglikelihood}
\end{equation}
where $B_{\epsilon}(p)=B(p)/\epsilon(p)$ is the efficiency-corrected background shape. 
The shape of the background in data is modeled by the background events in the signal region derived from the inclusive MC samples. 
The comparisons of the $M_{K^+\pi^+}$, $M_{K^+\pi^-}$ and $M_{\pi^+\pi^-}$ distributions of events outside the $D_s^+$ mass signal region between data and MC simulation validate the description from the inclusive MC samples. 
We have also examined the distributions of the background events in the inclusive MC samples inside and outside the $D_s^+$ mass signal region. Generally, they are compatible with each other within statistical uncertainties. The background shape $B(p)$ is modeled using RooNDKeysPDF~\cite{RooNDKeysPDF}, which is a kernel estimation method~\cite{CRANMER2001198} implemented in RooFit~\cite{RooNDKeysPDF} to model the distribution of an input dataset as a superposition of Gaussian kernels.

In the numerator of Eq.~\ref{pwa:pdf}, the $\epsilon(p)$ and $R_3(p)$ terms are independent of the fitted variables, so they are regarded as constants in the fit. As a consequence, the log-likelihood becomes
\begin{equation}
  \ln{\mathcal{L}} = \sum\limits_{i=1}^{3} \sum\limits_{k=1}^{N_{D,i}}{\rm ln}\Bigg[ \omega^{i}\frac{|\mathcal{M}(p^k)|^2}{\int \epsilon(p^k)|\mathcal{M}(p^k)|^2R_3(p^k)dp^k}+(1-\omega^{i})\frac{B_{\epsilon}(p^k)}{\int{\epsilon(p^k) B_\epsilon(p^k)R_3(p^k)}dp^k}\Bigg].
  \label{likelihoodfinal}
\end{equation}

The normalization integrals of signal and background are evaluated by MC integration,
\begin{equation}
\begin{aligned}
&\int \epsilon(p)|\mathcal{M}(p)|^2R_3(p)\,dp \approx \frac{1}{N_{\rm MC}}\sum^{N_{\rm MC}}_{k=1}\frac{|\mathcal{M}(p^{k})|^2}{|\mathcal{M}^{\rm gen}(p^{k})|^2},\\
&\int \epsilon(p)B_\epsilon(p)R_3(p)dp \approx \frac{1}{N_{\rm MC}}\sum^{N_{\rm MC}}_{k=1}\frac{B_\epsilon(p^{k})}{|\mathcal{M}^{\rm gen}(p^{k})|^2},
\end{aligned}
\end{equation}
where $k$ is the index of the $k^{\rm th}$ event of the MC sample and $N_{\rm MC}$ is the number of the selected MC events. 
The $M^{\rm gen}(p)$ is the signal PDF used to generate the MC samples in MC integration.

Tracking and PID differences between data and MC simulation are corrected for by multiplying the weight of the MC event by a factor $\gamma_{\epsilon}$, which is calculated as
\begin{equation}
  \gamma_{\epsilon}(p) = \prod_{n} \frac{\epsilon_{n,\rm data}(p)}{\epsilon_{n,\rm MC}(p)},
  \label{pwa:gamma}
\end{equation}
where $n$ refers to tracking or PID, $\epsilon_{n,\rm data}(p)$ and $\epsilon_{n,\rm MC}(p)$ are the tracking or PID efficiency as a function of the momenta of the daughter particles for data and MC simulation, respectively. {{The tracking and PID efficiencies are studied using clean samples of $e^+e^- \to  K^+K^-K^+K^-$, $e^+e^- \to K^+K^-\pi^+\pi^-$, $e^+e^- \to K^+K^-\pi^+\pi^-\pi^0$, $e^+e^- \to  \pi^+\pi^-\pi^+\pi^-$ and $e^+e^- \to  \pi^+\pi^-\pi^+\pi^-\pi^0$ processes.}} 
By weighting each signal MC event with $\gamma_{\epsilon}$, the MC integration is given by
\begin{equation}
  \int \epsilon(p)|\mathcal M(p)|^2R_3(p)dp \approx \frac{1}{N_{\rm MC}}\sum^{N_{\rm MC}}_{k=1}\frac{\gamma_{\epsilon}(p^{k_{\rm MC}})|\mathcal{M}(p^{k})|^2}{|\mathcal{M}^{\rm gen}(p^{k})|^2}.\label{likelihood3}
\end{equation}

\subsubsection{Propagator}
\label{Propagator}

The intermediate resonances $f_0(1370)$, $K^*(892)$ and $K^*(1410)^0$ are parameterized with the relativistic Breit-Wigner (RBW) formulas,
\begin{equation}
\begin{aligned}
		&P(m) = \frac{1}{(m^2_0-m^2)-im_0\Gamma(m)}, \\ 
		&\Gamma(m)=\Gamma_0\left(\frac{q}{q_0}\right)^{2L+1}\Big(\frac{m_0}{m}\Big)\left(\frac{X_L(q)}{X_L(q_0)}\right)^2,  \label{propagator}
\end{aligned}
\end{equation}
where $m^2$ is the invariant mass squared of the daughter particles of the intermediate resonances, $m_0$ and $\Gamma_0$ are the mass and width of the intermediate resonance, which are fixed to 1350 ${\rm MeV/}c^2$ and 265 MeV~\cite{BES:2004twe}, respectively, for $f_0(1370)$, and to the PDG values for the other resonances. In a process $a \to bc$, the variable $q$ is defined as
\begin{equation}
q = \sqrt{\frac{(s_a+s_b-s_c)^2}{4s_a}-s_b}, \label{q2}
\end{equation}
where $s_a, s_b,$ and $s_c$ are the invariant-mass squared of particles $a,\ b$ and $c$, respectively. The value of $q_0$ in Eq.~\ref{propagator} is that of $q$ when $s_a = m^2_a$, where $m_a$ is the mass of particle $a$.

The $\rho^0$ and $\rho(1450)$ mesons are parameterized as the Gounaris-Sakurai (GS) line shape~\cite{Gounaris:1968mw}, which is given by
\begin{equation}
P_{\rm GS}(m)=\frac{1+d\frac{\Gamma_0}{m_0}}{(m^2_0-m^2)+f(m)-im_0\Gamma(m)},
\end{equation}
where
\begin{equation}
	f(m)=\Gamma_0\frac{m^2_0}{q^3_0}\left[q^2(h(m)-h(m_0))+(m^2_0-m^2)q^2_0\frac{dh}{d(m^2)}\Big|_{m^2=m^2_0}\right]
\end{equation}
and the function $h(m)$ is defined as
\begin{equation}
h(m)=\frac{2}{\pi}\frac{q}{m}{\rm ln}\left(\frac{m+2q}{2m_{\pi}}\right),
\end{equation}
with
\begin{equation}
\frac{dh}{d(m^2)}\Big|_{m^2=m_0^2} = h(m_0)[(8q^2_0)^{-1}-(2m^2_0)^{-1}]+(2\pi m^2_0)^{-1},
\end{equation}
where $m_{\pi}$ is the mass of $\pi$, and the normalization condition at $P_{\rm GS}(0)$ fixes the parameter $d=\frac{f(0)}{\Gamma_0m_0}$. It is found to be
\begin{equation}
d=\frac{3}{\pi}\frac{m^2_{\pi}}{q^2_0}ln\left(\frac{m_0+2q_0}{2m_{\pi}}\right)+\frac{m_0}{2\pi q_0}-\frac{m^2_{\pi}m_0}{\pi q^3_0}.
\end{equation}

The $f_0(980)$ is parameterized with the Flatt\'e formula~\cite{BES:2004twe}:
\begin{equation}
	P_{f_0(980)}=\frac{1}{M_{f_0(980)}^2-m^2-i(g_{\pi\pi}\rho_{\pi\pi}(m^2)+g_{K\bar{K}}\rho_{K\bar{K}}(m^2))},
\end{equation}
where $g_{\pi\pi,K\bar{K}}$ are the constants coupling to individual final states. The parameters are fixed to be $g_{\pi\pi}=(0.165{ \pm 0.010 \pm 0.015)} {\rm GeV^2}/c^4$, $g_{K\bar{K}}=(4.21{ \pm 0.25 \pm 0.21)}g_{\pi\pi}$ and $M = 965$ MeV/$c^2$, as reported in Ref.~\cite{BES:2004twe}. The Lorentz invariant PHSP factors $\rho_{\pi\pi}(s)$ and $\rho_{K\bar{K}}(s)$ are given by
\begin{equation}
\begin{aligned}
	\rho_{\pi\pi}=\frac{2}{3}\sqrt{1-\frac{4m^2_{\pi^\pm}}{m^2}}+\frac{1}{3}\sqrt{1-\frac{4m^2_{\pi^0}}{m^2}},\\
	\rho_{K\bar{K}}=\frac{1}{2}\sqrt{1-\frac{4m^2_{K^\pm}}{m^2}}+\frac{1}{2}\sqrt{1-\frac{4m^2_{K^0}}{m^2}}.
\end{aligned}
\end{equation}

The resonance $f_0(500)$ is parameterized with the formula given in Ref.~\cite{BUGG199659}:
\begin{equation}
\begin{aligned}
	P_{f_0(500)}=\frac{1}{m_0^2-m^2-im_0\Gamma_{\rm tot}(m)},
\end{aligned}
\end{equation}
where $\Gamma_{\rm tot}(m)$ is decomposed into two parts:
\begin{equation}
\begin{aligned}
	\Gamma_{\rm tot}(m)=g_{\pi\pi}\frac{\rho_{\pi\pi}(m)}{\rho_{\pi\pi}(m_0)}+g_{4\pi}\frac{\rho_{4\pi}(m)}{\rho_{4\pi}(m_0)}
\end{aligned}
\end{equation}
and
\begin{equation}
\begin{aligned}
	g_{\pi\pi}=(b_1+b_2m^2)
   \left( \frac{m^2-m_\pi^2/2}{m_0^2-m_\pi^2/2} \right) \; e^{(m_0^2-m^2)/a},
\end{aligned}
\end{equation}
where $\rho_{\pi\pi}$ is the PHSP of the $\pi^+\pi^-$ system and $\rho_{4\pi}$ is the PHSP of the $4\pi$ system and is approximated by
\begin{equation}
\begin{aligned}
	\rho_{4\pi}  = \frac{\sqrt{\left(1-\frac{1-16m_\pi^2}{m^2}\right)}}{1+e^{3.5(2.8-m^2)}},
\end{aligned}
\end{equation}
with the parameters fixed to the values given in Ref.~\cite{Pelaez:2015qba}.

The $K^*_0(1430)$ is parameterized with the Flatt\'e formula:
\begin{equation}
	P_{K^*_0(1430)}=\frac{1}{M_{K^*_0(1430)}^2-m^2-i(g_{K\pi}\rho_{K\pi}(m^2)+g_{\eta\prime K}\rho_{\eta\prime K}(m^2))},
\end{equation}
where $\rho_{K\pi}(m^2)$ and $\rho_{\eta\prime K}(m^2)$ are the Lorentz invariant PHSP factors, and $g_{{K^+\pi^-},{\eta\prime K}}$ are the constants coupling to individual final states. The parameters of the $K^*_0(1430)$ are fixed to 
$M = 1471.2$ MeV/c$^2$,  $g_{K^+\pi^-} = 546.8$ MeV/c$^2$, and 
$g_{\eta\prime K} = 230$ MeV/c$^2$, from CLEO~\cite{CLEO:2008jus}.

{{The $K\pi$ S-wave modeled by the LASS parameterization~\cite{Aston:1987ir} is described by a $K^*_0(1430)$ Breit-Wigner together with an effective range non-resonant component with a phase shift. It is given by
\begin{equation}
A(m) = F{\rm sin}\delta_F e^{i\delta_F} + R{\rm sin}\delta_R e^{i\delta_R}e^{i2\delta_F},
\end{equation}
with
\begin{equation}
\begin{aligned}
&\delta_F = \phi_F + {\rm cot}^{-1}\Big[\frac{1}{aq}+\frac{rq}{2}\Big],\\
&\delta_R = \phi_R + {\rm tan}^{-1}\Big[\frac{M\Gamma(m_{K\pi})}{M^2-m^2_{K\pi}}\Big],
\end{aligned}
\end{equation}
where the parameters $F(\phi_F)$ and $R(\phi_R)$ are the magnitudes (phases) for non-resonant state and resonance terms, respectively. The parameters $a$ and $r$ are the scattering length and effective interaction length, respectively. We fix these parameters ($M, \Gamma, F, \phi_F, R, \phi_R, a, r$) to the results obtained from the amplitude analysis to a sample of $D^0 \to K_S^0\pi^+\pi^-$ by the BABAR and Belle experiments~\cite{PhysRevD.98.112012}; these parameters are summarised in Table~\ref{LASSpa}.

\begin{table}[htbp]

  \begin{center}
    \begin{tabular}{| l c |}
      \hline
      $M\ ({\rm GeV}/c^2)$ &1.441 $\pm$ 0.002\\
      $\Gamma\ ({\rm GeV})$ &0.193 $\pm$ 0.004\\
      $F$ &0.96 $\pm$ 0.07\\
      $\phi_F\ (^\circ)$ &0.1 $\pm$ 0.3\\
      $R$ &{1(fixed)}\\
      $\phi_R\ (^\circ)$ &-109.7 $\pm$ 2.6\\
      $a\ ({\rm GeV}/c)^{-1}$ &0.113 $\pm$ 0.006\\
      $r\ ({\rm GeV}/c)^{-1}$ &-33.8 $\pm$ 1.8\\
      \hline
    \end{tabular}
  \end{center}
  \caption{The $K\pi$ S-wave parameters, obtained from the amplitude analysis of $D^0 \to K_S^0\pi^+\pi^-$ by the BABAR and Belle experiments~\cite{PhysRevD.98.112012}. Uncertainties are statistical only.}
  \label{LASSpa}
\end{table}

}}

\subsubsection{Spin factors}
\label{Spinfactors}
The spin-projection operators are defined as~\cite{covariant-tensors}
\begin{equation}
\begin{aligned}
&P^0(a) = 1,&(S\ \rm wave)\\
&P^{(1)}_{\mu\nu}(a)=-g_{\mu\nu}+\frac{p_{a\mu}p_{a\nu}}{p^2_a},&(P\ \rm wave)\\
&P^{(2)}_{\mu_1\mu_2\nu_1\nu_2}(a)=\frac{1}{2}\left(P^{(1)}_{\mu_1\nu_1}P^{(1)}_{\mu_2\nu_2}+P^{(1)}_{\mu_1\nu_2}P^{(1)}_{\mu_2\nu_1}\right)-\frac{1}{3}P^{(1)}_{\mu_1\mu_2}P^{(1)}_{\nu_1\nu_2}.&(D\ \rm wave)
\end{aligned}
\end{equation}
The $p_a,\ p_b$, and $p_c$ variables are the momenta of particles $a,\ b$ and $c$, respectively, and $r_a = p_b - p_c$. The covariant tensors are given by
\begin{equation}
\begin{aligned}
&\tilde{t}^{(0)}(a) = 1,&(S\ {\rm wave})\\
&\tilde{t}^{(1)}_\mu(a) = -P^{(1)}_{\mu\nu}(a)r_a^{\nu},&(P\  \rm wave)\\
&\tilde{t}^{(2)}_{\mu\nu}(a) = P^{(2)}_{\mu\mu_1\nu\nu_2}(a)r_a^{\mu_1}r_a^{\nu_1}.&(D\ \rm wave)
\end{aligned}
\end{equation}
The spin factors for $S,P$, and $D$ wave decays are
\begin{equation}
\begin{aligned}
&S_n = 1,&(S\ \rm wave)\\
&S_n = \tilde{T}^{(1)\mu}(D_s)\tilde{t}^{(1)}_\mu(a),&(P\ \rm wave)\\
&S_n = \tilde{T}^{(2)\mu\nu}(D_s)\tilde{t}^{(2)}_{\mu\nu}(a),&(D\ \rm wave)
\end{aligned}
\end{equation}
where $\tilde{T}$ has the same definition as $\tilde{t}$. The tensor describing the $D_s^+$ decays is denoted by $\tilde{T}$ and that of $a$ decays is denoted by $\tilde{t}$.

\subsubsection{Blatt-Weisskopf barriers}
\label{Blatt-Weisskopfbarriers}
For a decay process $a \to b\ c$, the Blatt-Weisskopf barriers {{factors}}~\cite{PhysRevD.104.012016} depend on the angular momenta $L$ 
and the momentum $q$ of the final-state particle $b$ or $c$ in the rest system of $a$. 
They are taken as
\begin{equation}
\begin{aligned}
  &X_{L=0}(q)=1,\\
  &X_{L=1}(q)=\sqrt{\frac{z_0^2+1}{z^2+1}},\\
  &X_{L=2}(q)=\sqrt{\frac{z_0^4+3z_0^2+9}{z^4+3z^2+9}}, \label{xl}
\end{aligned}
\end{equation}
where $z = qR$ and $z_0 = q_0R$ with $q_0$ defined in Sec.~\ref{Propagator}. The effective radius of barrier $R$ is fixed to be 3.0 GeV$^{-1}$ for the intermediate resonances and 5.0 GeV$^{-1}$ for the $D_s^+$ meson.

\subsection{Fit results}
\begin{figure}[htp]
          \centering
          \includegraphics[width=7.cm]{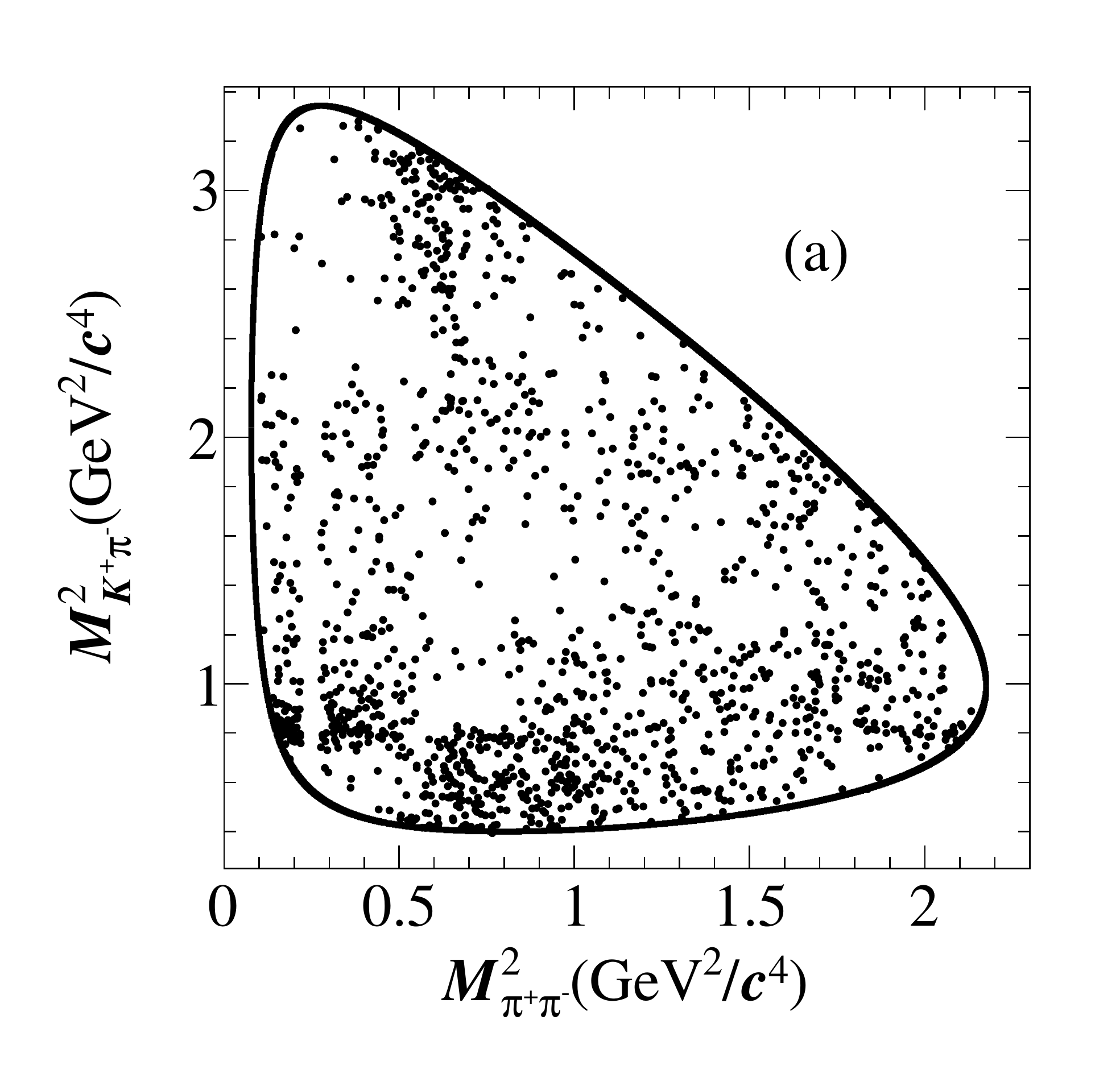}
          \includegraphics[width=7.cm]{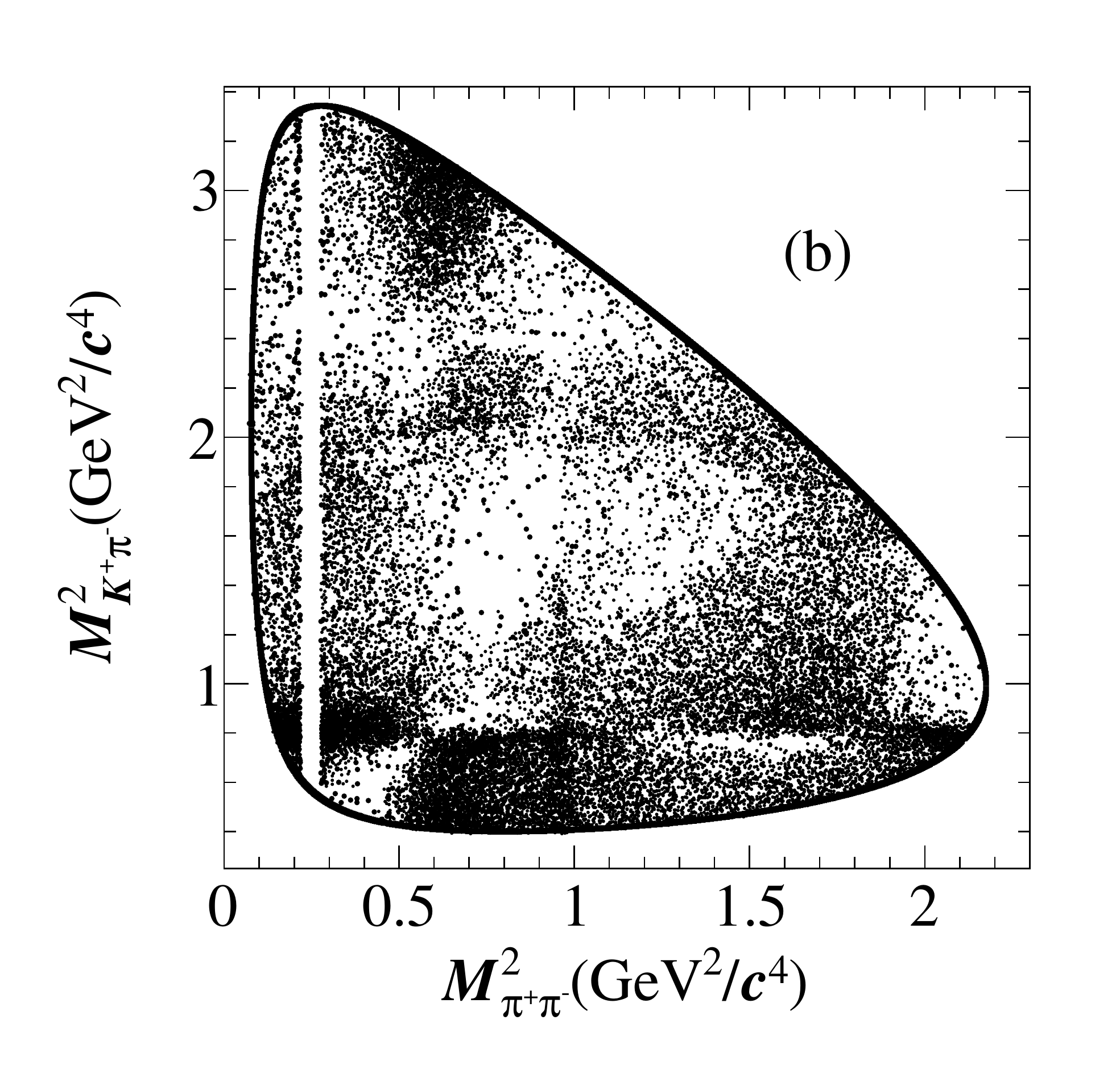}
          \caption{The Dalitz plots of $M_{K^+\pi^-}^2$ versus $M_{\pi^+\pi^-}^2$ of the selected DT candidates from (a) the data sample and (b) the signal MC sample generated based on the amplitude analysis results at $\sqrt{s} = 4.178-4.226 {\rm \ GeV}$. The black lines indicate the physical border.}
    \label{dalitz}
\end{figure}

Figures~\ref{dalitz}(a) and~\ref{dalitz}(b) show the Dalitz plots of $M^2_{K^+\pi^-}$ versus $M^2_{\pi^+\pi^-}$ of the selected DT candidates from the data samples and the signal MC samples generated based on the results of the amplitude analysis, respectively. In the fit, the magnitude and phase of the reference amplitude $D_s^+ \to K^+\rho^0$ are fixed to 1.0 and 0.0, respectively, while those of the other amplitudes are left floating. The $\omega^i$ values are fixed to the purities given in Sec.~\ref{AASelection}. 

In addition to the dominant amplitudes of $D^+_s \to K^+\rho^0$ and $D^+_s \to K^*(892)^0\pi^+$, we have also tested other possible intermediate resonances, including $K^*(1410)^0$, $K_0^*(1430)^0$, $f_0(500)$, $f_0(980)$, $f_0(1370)$, $\rho(1450)^0$, $(K^+\pi^-)_{\rm S-wave}$ (using LASS parameterization~\cite{Aston:1987ir} or $K$-matrix~\cite{Anisovich:1997qp}), etc.  Finally, the amplitudes of $D_s^{+}\to K^+\rho^0$, $D_s^{+}\to K^+\rho(1450)^0$, $D_s^{+}\to K^+f_0(500)$, $D_s^{+}\to K^+f_0(980)$, $D_s^{+}\to K^+f_0(1370)$, $D_s^{+}\to K^*(892)^0\pi^+$, $D_s^{+}\to K^*(1410)^0\pi^+$, and $D_s^{+}\to K^*_0(1430)^0\pi^+$, which have statistical significances greater than five standard deviations, are retained in the nominal fit. The statistical significances are determined from the changes in log-likelihood and the numbers of degrees of freedom (NDOF) between the fits with a given amplitude included or excluded.

The PHSP MC truth information without detector acceptance and resolution effects is used to calculate the fit fractions (FFs) for individual amplitudes. The FF for the $n^{\rm th}$ amplitude is defined as
\begin{eqnarray}\begin{aligned}
	{\rm FF}_{n} = \frac{\sum^{N_{\rm gen}} \left|c_n A_{n}\right|^{2}}{\sum^{N_{\rm gen}} \left|\mathcal M\right|^{2}}\,, \label{Fit-Fraction-Definition}
\end{aligned}\end{eqnarray}
where $N_{\rm gen}$ is the number of PHSP signal MC events at generator level. {{Interference IN between the $n^{\rm th}$ and $n^{\rm \prime th}$ amplitudes is defined as
\begin{equation}
    {\rm IN}_{nn^\prime}
    = \frac{\sum^{N_{\rm gen}}2{\rm Re}[c_nc^*_{n^\prime}\mathcal{A}_n\mathcal{A}^{*}_{n^\prime}]}{\sum^{N_{\rm gen}}|\mathcal{M}|^2},
\end{equation}
}}
The statistical uncertainties of FFs are obtained by randomly perturbing the fit parameters according to their uncertainties and covariance matrix and re-evaluating FFs.  A Gaussian function is fit to the resulting distribution for each FF and the fitted width is taken as its statistical uncertainty.  

The phases, FFs and statistical significances for various amplitudes are listed in Table~\ref{tab:signi}. {{The interference between amplitudes is listed in Table~\ref{interference}.}} The statistical significances for amplitudes tested but not included in the nominal fit are listed in Table~\ref{tab:tested_amplitude}.
 
\begin{figure}[htp]
          \centering
          \includegraphics[width=7.cm]{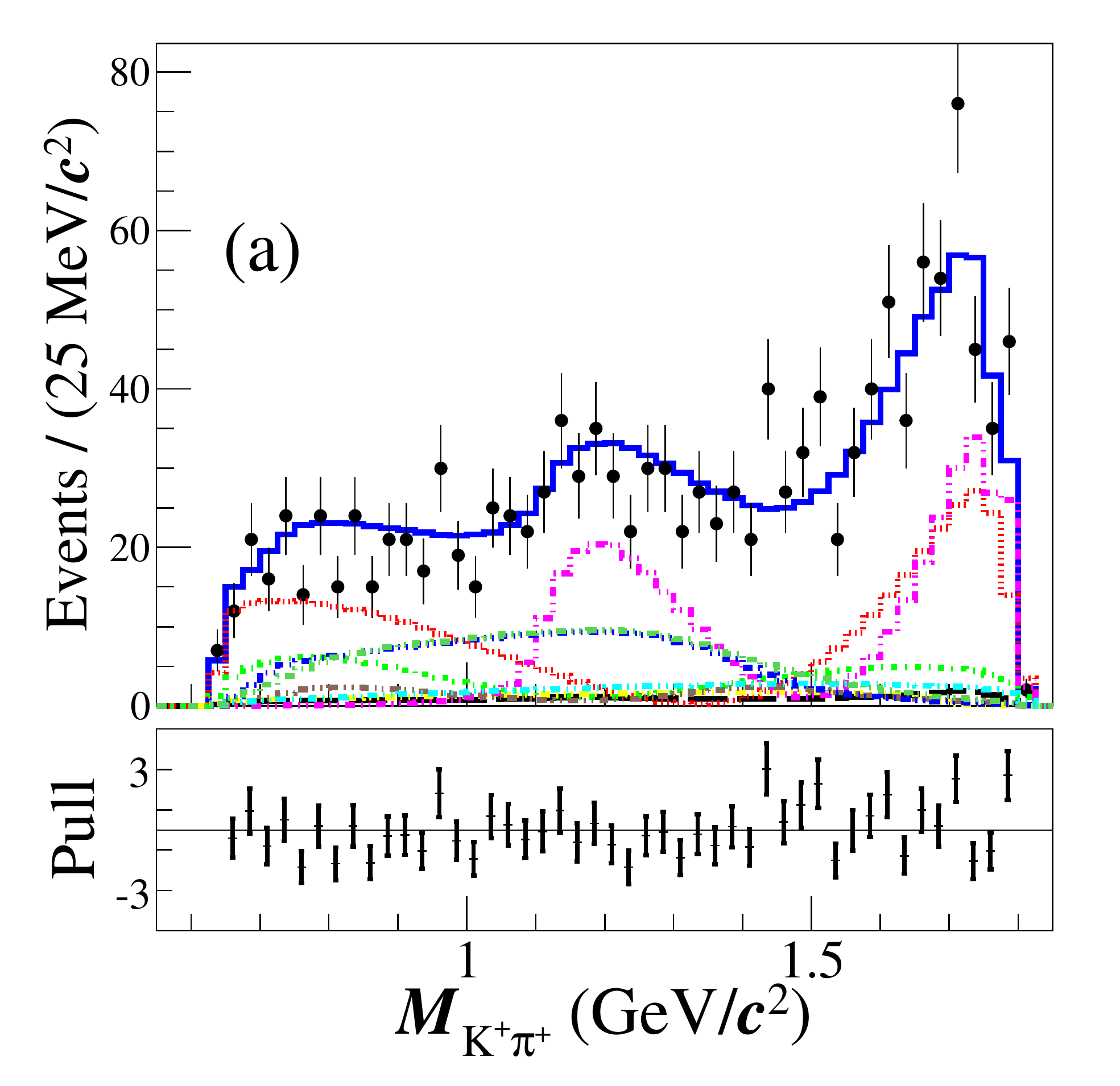}
          \includegraphics[width=7.cm]{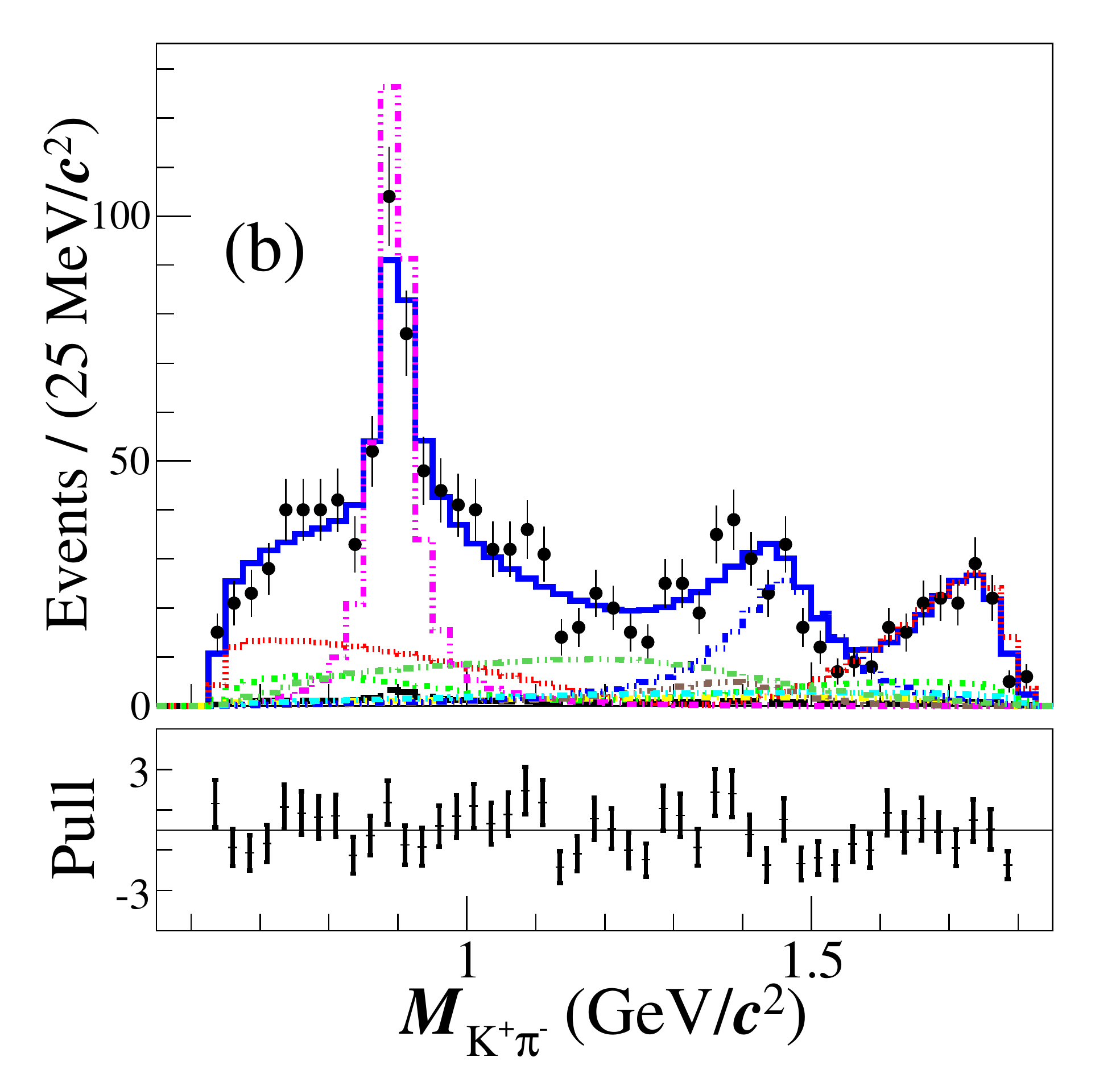}
          \includegraphics[width=7.cm]{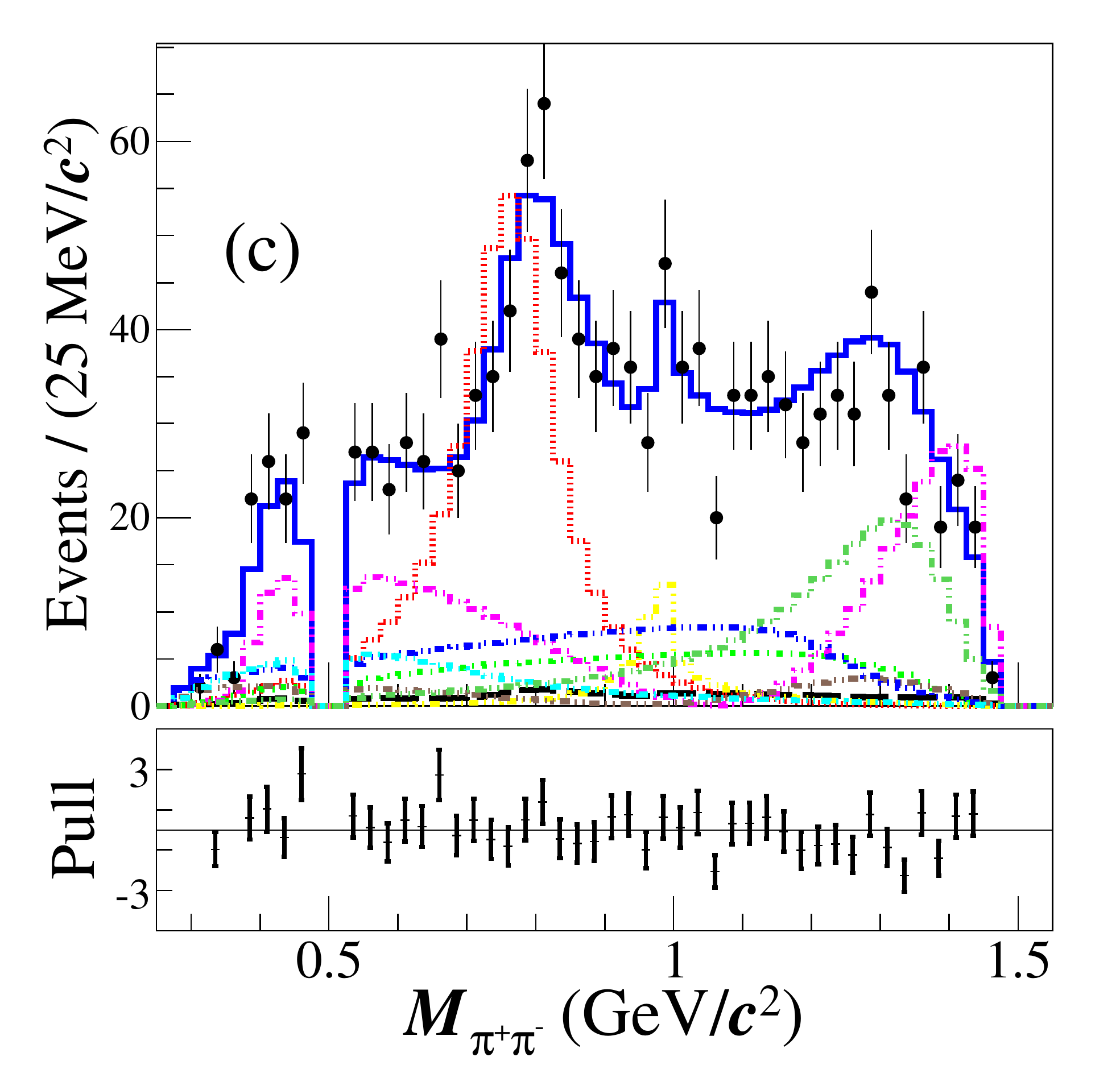}
          \includegraphics[width=7.cm]{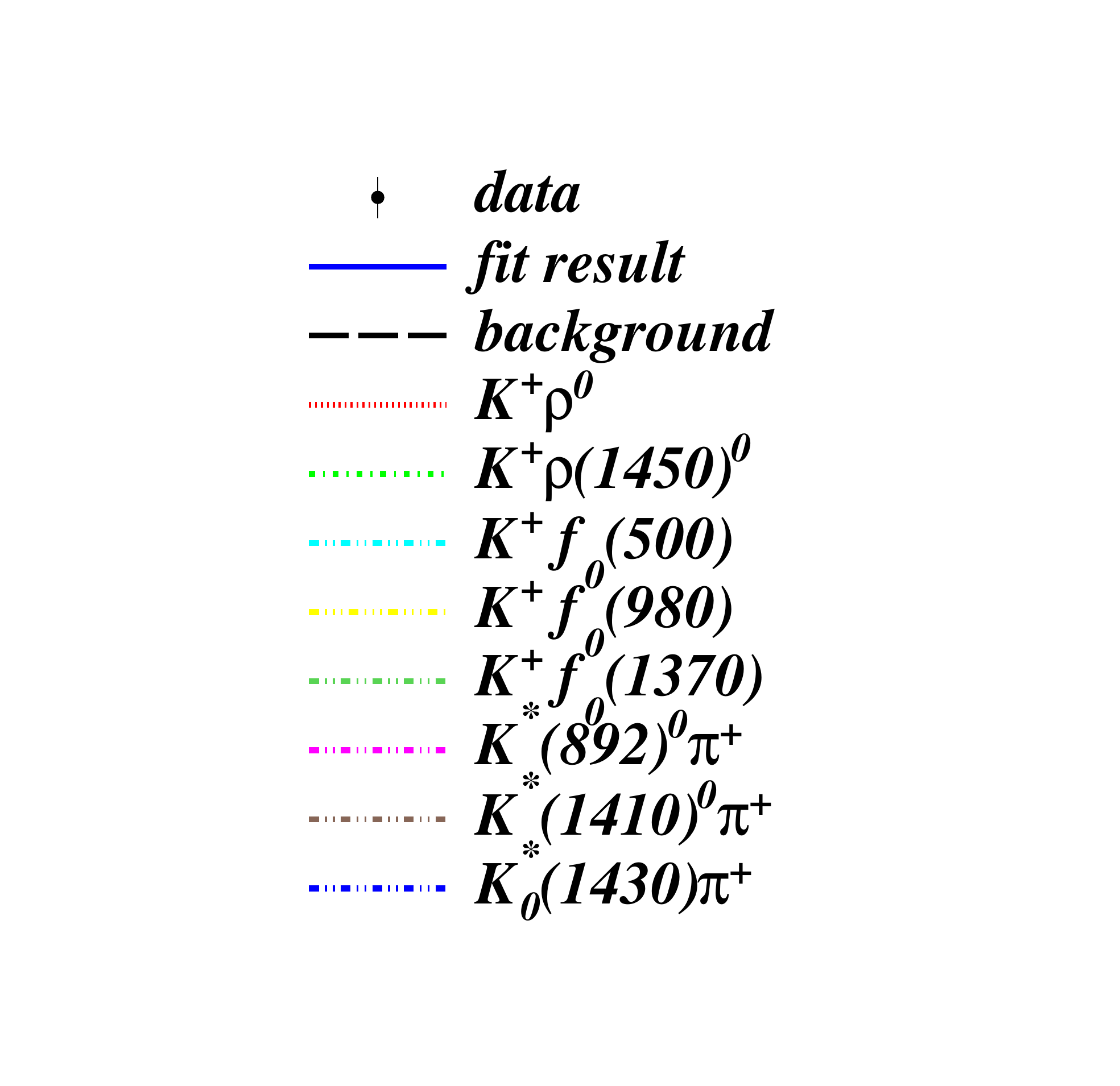}
          \caption{The projections on (a) $M_{K^+\pi^+}$, (b) $M_{K^+\pi^-}$, and (c) $M_{\pi^+\pi^-}$ of the nominal fit. The data samples at $\sqrt{s} = 4.178-4.226\ \rm{GeV}$ are represented by points with error bars, the fit results by the solid blue lines, and the background estimated from the inclusive MC samples by the black dashed lines. Colored curves show the components of the fit model. Due to interference effects, the total is not necessarily equal to the linear sum of the components. Pull projections are shown beneath each distribution; if there are less than 10 events in a bin, that bin is merged with the next bin until the number of events is larger than or equal to 10.}
    \label{pwa:proji}
\end{figure}

The mass projections of the nominal fit for the amplitude analysis are shown in Fig.~\ref{pwa:proji}. Their systematic uncertainties will be discussed in next section. The sum of the FFs is not unity due to interferences among amplitudes.

\begin{table}[htbp]\small

	\centering
	\begin{tabular}{l r@{ $\pm$ }c@{ $\pm$ }c r@{ $\pm$ }c@{ $\pm$ }c c}
	\hline
  Amplitude &\multicolumn{3}{c}{Phase $\phi_n$ (rad)} &\multicolumn{3}{c}{FF(\%)} &Statistical significance($\sigma$)\\
	\hline
		 $D_s^{+}\to K^+\rho^0$  &\multicolumn{3}{c}{0.0 (fixed)} &{{32.1}} &{{3.7}}  &3.7 &>10 \\
		 $D_s^{+}\to K^+\rho(1450)^0$  &{{2.74}} & 0.14 & {0.24} &{{13.1}} &{{3.1}}  &{{2.9}} & >10\\
		 $D_s^{+}\to K^+f_0(500)$      &{{1.01}} & 0.17 & {{0.28}} & {{7.2}} &{{2.1}}  &{{4.4}} & 6.8\\
		 $D_s^{+}\to K^+f_0(980)$      &{{5.05}} & 0.15 & {{0.17}} &{{4.5}} &1.3  &{{1.2}}& 6.9\\
		 $D_s^{+}\to K^+f_0(1370)$     &{{6.04}} & 0.14 & {0.26} &19.9 &{{2.9}}  &{{9.3}}& >10\\ 
		 $D_s^{+}\to K^*(892)^0\pi^+$  &3.03 &{{0.08}} &{0.04} &{{30.2}} &{{1.8}}  &{{2.0}}& >10\\
		 $D_s^{+}\to K^*(1410)^0\pi^+$ &{{5.60}} & 0.14 & {0.09} &{{4.5}} &{{2.1}}  &{{2.5}} &5.2\\
		 $D_s^{+}\to K^*_0(1430)^0\pi^+$ &{{1.90}} & 0.19 & {{0.20}} &{{18.5}} &2.5  &{{2.6}}& 8.6\\
    \hline
    \end{tabular}
	\caption{The phases, FFs and statistical significances for various amplitudes in the nominal fit. The first and second uncertainties of the phases and FFs are statistical and systematic, respectively. The total FF is {{130.1}}\%.}
	\label{tab:signi}
\end{table}

\begin{table}[htbp]\small
	\centering
	\resizebox{\textwidth}{!}{
	\begin{tabular}{c|r@{ $\pm$ }c r@{ $\pm$ }c r@{ $\pm$ }c r@{ $\pm$ }c r@{ $\pm$ }c r@{ $\pm$ }c r@{ $\pm$ }c r@{ $\pm$ }c}
	\hline
		 &\multicolumn{2}{c}{I} &\multicolumn{2}{c}{II} &\multicolumn{2}{c}{III} &\multicolumn{2}{c}{IV} &\multicolumn{2}{c}{V} &\multicolumn{2}{c}{VI} &\multicolumn{2}{c}{VII} &\multicolumn{2}{c}{VIII} \\
	\hline
		 I &{{32.1}} &{{3.7}} &1.8 & 3.0 &0.0 & 0.0 &0.0 & 0.0 &0.0 & 0.0 &-6.5 & 0.4 &1.1 & 0.8 &-7.2 & 0.6 \\
		 II &\multicolumn{2}{c}{} &{{13.1}} &{{3.1}} &0.0 & 0.0 &0.0 & 0.0 &0.0 & 0.0 &-5.4 & 1.3 &-4.8 & 1.0 &3.9 & 2.7\\
		 III &\multicolumn{2}{c}{} &\multicolumn{2}{c}{} &{{7.2}} &{{2.1}} &-4.8 & 1.4 &-6.1 & 2.5 &2.9 & 0.5 &-1.3 & 0.8 &4.5 & 1.1\\
		 IV &\multicolumn{2}{c}{} &\multicolumn{2}{c}{} &\multicolumn{2}{c}{} &{{4.5}} &1.3 &10.7 & 1.6 &-2.3 & 0.4 &0.3 & 0.4 &-3.9 & 0.7\\
		 V &\multicolumn{2}{c}{} &\multicolumn{2}{c}{} &\multicolumn{2}{c}{} &\multicolumn{2}{c}{} &19.9 &{{2.9}} &-8.4 & 0.8 &0.5 & 1.1 &-9.4 & 2.0\\
		 VI  &\multicolumn{2}{c}{} &\multicolumn{2}{c}{} &\multicolumn{2}{c}{} &\multicolumn{2}{c}{} &\multicolumn{2}{c}{} &{{30.2}} &{{1.8}} &4.4 & 0.9 &0.0 & 0.0\\
		 VII &\multicolumn{2}{c}{} &\multicolumn{2}{c}{} &\multicolumn{2}{c}{} &\multicolumn{2}{c}{} &\multicolumn{2}{c}{} &\multicolumn{2}{c}{} &{{4.5}} &{{2.1}} &0.0 & 0.0\\
		 VIII &\multicolumn{2}{c}{} &\multicolumn{2}{c}{} &\multicolumn{2}{c}{} &\multicolumn{2}{c}{} &\multicolumn{2}{c}{} &\multicolumn{2}{c}{} &\multicolumn{2}{c}{} &{{18.5}} & 2.5 \\
    \hline
    \end{tabular}
    }
    \caption{Interference between amplitudes, in unit of \% of total amplitude. I denotes $D_s^{+}\to K^+\rho^0$, II $D_s^{+}\to K^+\rho(1450)^0$, III $D_s^{+}\to K^+f_0(500)$, IV $D_s^{+}\to K^+f_0(980)$, V $D_s^{+}\to K^+f_0(1370)$, VI $D_s^{+}\to K^*(892)^0\pi^+$, VII $D_s^{+}\to K^*(1410)^0\pi^+$,  and VIII $D_s^{+}\to K^*_0(1430)^0\pi^+$. The uncertainties are statistical only.}
		\label{interference}
\end{table}

\begin{table}[htbp]

  \begin{center}
    \begin{tabular}{lc}
      \hline
      Amplitude                                          &Statistical significance($\sigma$)\\
      \hline
      $D_s^{+}\to K^+f_2(1270)$  &2.3\\
      $D_s^{+}\to K^+f_0(1500)$ &3.1\\
      $D_s^{+}\to K^+f_2^\prime(1525)$ &2.3\\
      $D_s^{+}\to K^*(1680)^0\pi^+$ &3.3\\
      $D_s^{+}\to (K^+\pi^-)_{\rm S-wave}\pi^+$ &<1\\
      \hline
    \end{tabular}
  \end{center}
  \caption{Statistical significances for amplitudes tested, but not included in the nominal fit.}
  \label{tab:tested_amplitude}
\end{table}

\subsection{Systematic uncertainties for the amplitude analysis}
\label{sec:PWA-Sys}
The systematic uncertainties for the amplitude analysis are summarized
in Table~\ref{pwa:sys}, and are described below.
\begin{itemize}
\item[\lowercase\expandafter{\romannumeral1}]
Fixed parameters in the amplitudes. The masses and widths of $K^*(892)$ and $K^*(1410)^0$ are shifted by their corresponding uncertainties~\cite{PDG}. The mass and width of $f_0(1370)$ are shifted according to the uncertainties from Ref.~\cite{BES:2004twe}. The masses and coupling constants of the $f_0(980)$ and $K^*(1430)$ Flatt\'e formulas are varied according to Ref.~\cite{BES:2004twe} and Ref.~\cite{CLEO:2008jus}, respectively. The uncertainties of the lineshapes of $\rho^+$ and $\rho(1450)^0$ are estimated by replacing the GS with the RBW formula. The uncertainties of the lineshape of $f_0(500)$ are estimated by replacing the propagator with a RBW function with the mass and width fixed at 526 MeV and 534 MeV, respectively~\cite{Pelaez:2015qba}.
The changes of the phases $\phi$ and FFs are assigned as the associated systematic uncertainties.

\item[\lowercase\expandafter{\romannumeral2}]
$R$ values. The estimation of the systematic uncertainty associated with the $R$ parameters in the Blatt-Weisskopf factors is performed by repeating the fit procedure after varying the radii of the intermediate states and $D_s^+$ mesons by $\pm 1$ GeV$^{-1}$.

\item[\lowercase\expandafter{\romannumeral3}]
Fit bias. 
An ensemble of 600 signal MC samples is generated according to the results of the amplitude analysis. The pull distribution, which is supposed to be a normal distribution, is used to validate the fit performance. The fitted pull values for FFs of $D_s^+ \to K^+\rho^0$, $D_s^+ \to K^+f_0(980)$ and $D_s^+ \to K^*_0(1430)^0\pi^+$ and the fitted pull values for phases of $D_s^+ \to K^+f_0(500)$, $D_s^+ \to K^+f_0(980)$ and $D_s^+ \to K^*(892)^0\pi^+$ deviate from zero by more than { three, but less than five, standard deviations. Hence, the differences between input values and average fit results are taken as the systematic uncertainties.}

\item[\lowercase\expandafter{\romannumeral4}]
Background estimation.
The fractions of signal, i.e. $\omega^i$ in Eq.~\ref{loglikelihood}, are varied within their uncertainties and the largest difference from the fits is taken as the uncertainty from the background level. 
The uncertainty corresponding to the background shape is determined by replacing the input parameters (keeping $M^2_{K^+\pi^-}$ but replacing $M^2_{\pi^+\pi^-}$ with $M^2_{K^+\pi^+}$) and changing the smoothing parameters in RooNDKeysPdf~\cite{RooNDKeysPDF}.

\item[\lowercase\expandafter{\romannumeral5}]
Experimental effects.
The systematic uncertainty from knowledge of the $\gamma_{\epsilon}$ factors in 
Eq.~(\ref{likelihood3}), which correct for data-MC differences in tracking and PID efficiencies, 
is evaluated by performing the fit after varying the weights according to their uncertainties.

{{\item[\lowercase\expandafter{\romannumeral6}]
Insignificant amplitudes.
The intermediate resonances with statistical significances less than $5\sigma$ in Table~\ref{tab:tested_amplitude} are added to the model one by one. The largest variations from the nominal result are taken as the corresponding systematic uncertainties.}}

\end{itemize}

\begin{table}[h]
\renewcommand\arraystretch{1.25}
  \centering

	\begin{tabular}{lcccccccc}
    \hline
    \multirow{2}{*}{Amplitude}&\multicolumn{7}{c}{Source}\cr
		& & \lowercase\expandafter{\romannumeral1} &\lowercase\expandafter{\romannumeral2} &\lowercase\expandafter{\romannumeral3} &\lowercase\expandafter{\romannumeral4} &\lowercase\expandafter{\romannumeral5} 
		&{{\lowercase\expandafter{\romannumeral6}}}
		& Total   \\
	\hline
		$D_s^{+}\to K^+\rho^0$ &FF &1.10 &0.58 &0.05 &0.04 &0.01 &{{0.15}} &{{1.26}} \\ 
		\hline
		\multirow{2}{*}{$D_s^{+}\to K^+\rho(1450)^0$} &$\phi$ &{1.62} &0.63 &{{0.13}} &0.07 &0.06 &{{0.14}} &{{1.75}} \\ &FF &0.81 &0.25 &{{0.07}} &0.09 &0.00 &{{0.28}} &{{0.91}}\\
		\hline
		\multirow{2}{*}{$D_s^{+}\to K^+f_0(500)$} &$\phi$ &1.08 &0.22 &{{0.15}} &0.12 &0.00 &{{1.18}} &{{1.62}} \\ &FF &1.77 &0.43 &{{0.12}} &0.00 &0.33 &{{0.83}} &{{2.04}} \\
		\hline
		\multirow{2}{*}{$D_s^{+}\to K^+f_0(980)$} &$\phi$ &0.99 &0.13 &{{0.19}} &0.07 &0.00 &{{0.47}} &{{1.11}} \\ &FF &0.83 &0.17 &{{0.00}} &0.09 &0.02 &{{0.34}} &{{0.95}} \\
		\hline
		\multirow{2}{*}{$D_s^{+}\to K^+f_0(1370)$} &$\phi$ &1.82 &0.07 &{{0.25}} &0.07 &0.00 &{{0.43}} &{{1.87}} \\ &FF &0.97 &0.20 &{{0.07}} &0.01 &0.02 &{{3.03}} &{{3.19}} \\
		\hline
		\multirow{2}{*}{$D_s^{+}\to K^*(892)^0\pi^+$} &$\phi$ &0.45 &0.13 &0.05 &0.13 &0.00 &{{0.13}} &{{0.50}} \\ &FF &0.92 &0.32 &{{0.00}} &0.11 &0.02 &{{0.51}} &{{1.11}} \\
		\hline
		\multirow{2}{*}{$D_s^{+}\to K^*(1410)^0\pi^+$} &$\phi$ &0.56 &0.31 &{{0.14}} &0.07 &0.08 &{{0.07}} &{{0.67}} \\ &FF &0.79 &0.62 &{{0.03}} &0.10 &0.03 &{{0.67}} &{{1.21}} \\
		\hline
		\multirow{2}{*}{$D_s^{+}\to K^*_0(1430)^0\pi^+$} &$\phi$ &0.90 &0.20 &0.05 &0.10 &0.00&{{0.53}} &{{1.07}} \\ &FF &0.92 &0.27 &{{0.15}} &0.04 &0.03 &{{0.35}} &{{1.03}} \\
	\hline
	\end{tabular}
	  \caption{Systematic uncertainties on the $\phi$ and FF for each amplitude in units of the corresponding statistical uncertainty. The sources are:
    (\lowercase\expandafter{\romannumeral1}) fixed parameters in the amplitudes,
    (\lowercase\expandafter{\romannumeral2}) $R$ values,
    (\lowercase\expandafter{\romannumeral3}) fit bias,
	(\lowercase\expandafter{\romannumeral4}) background estimation,
	(\lowercase\expandafter{\romannumeral5}) experiment effects{{,
	(\lowercase\expandafter{\romannumeral6}) insignificant amplitudes}}.
}
	\label{pwa:sys}
\end{table}

\section{{{BF}} measurement}
\label{BFSelection}
The BF measurements are based on the following equations:
\begin{eqnarray}\begin{aligned}
  N_{\text{tag}}^{\text{ST}} = 2N_{D_{s}^{+}D_{s}^{-}}\mathcal{B}_{\text{tag}}\epsilon_{\text{tag}}^{\text{ST}}\,, \label{eq-ST}
\end{aligned}\end{eqnarray}
\begin{equation}
  N_{\text{tag,sig}}^{\text{DT}}=2N_{D_{s}^{+}D_{s}^{-}}\mathcal{B}_{\text{tag}}\mathcal{B}_{\text{sig}}\epsilon_{\text{tag,sig}}^{\text{DT}}\,,
  \label{eq-DT}
\end{equation}
where $N_{\text{tag}}^{\text{ST}}$ is the ST yield for the tag mode, $N_{\text{tag,sig}}^{\text{DT}}$ is the DT yield, $N_{D_{s}^{+}D_{s}^{-}}$ is the total number of $D_{s}^{*\pm}D_{s}^{\mp}$ pairs produced from the $e^{+}e^{-}$ collisions, $\mathcal{B}_{\text{tag}}$ and $\mathcal{B}_{\text{sig}}$ are the BFs of the tag and signal modes, respectively. The $\epsilon_{\text{tag}}^{\text{ST}}$ is the efficiency to reconstruct the tag mode alone and $\epsilon_{\text{tag,sig}}^{\text{DT}}$ is the efficiency to reconstruct both the tag and signal modes. 
In the case of more than one tag mode and energy group,
\begin{eqnarray}
\begin{aligned}
  \begin{array}{lr}
    N_{\text{total}}^{\text{DT}}=\Sigma_{\alpha, i}N_{\alpha,\text{sig},i}^{\text{DT}}   = \mathcal{B}_{\text{sig}}
 \Sigma_{\alpha, i}2N_{D_{s}^{+}D_{s}^{-},i}\mathcal{B}_{\alpha}\epsilon_{\alpha,\text{sig}, i}^{\text{DT}}\,,
  \end{array}
  \label{eq-DTtotal}
\end{aligned}
\end{eqnarray}
where $\alpha$ represents tag modes in the $i^{\rm th}$ energy group.
Solving for $\mathcal{B}_{\text{sig}}$, 
\begin{eqnarray}\begin{aligned}
  \mathcal{B}_{\text{sig}} =
  \frac{N_{\text{total}}^{\text{DT}}}{ \begin{matrix}\sum_{\alpha, i} N_{\alpha, i}^{\text{ST}}\epsilon^{\text{DT}}_{\alpha,\text{sig},i}/\epsilon_{\alpha,i}^{\text{ST}}\end{matrix}},
\end{aligned}\end{eqnarray}
where $N_{\alpha,i}^{\text{ST}}$ and $\epsilon_{\alpha,i}^{\text{ST}}$ are obtained from the data and inclusive MC samples, respectively. The $\epsilon_{\alpha,\text{sig},i}^{\text{DT}}$ is determined with signal MC samples in which $D_{s}^{+} \to K^{+}\pi^{+}\pi^{-}$ events are generated according to the baseline model of the amplitude analysis. 

\begin{figure*}[htp]
\begin{center}
    \includegraphics[width=6.0cm]{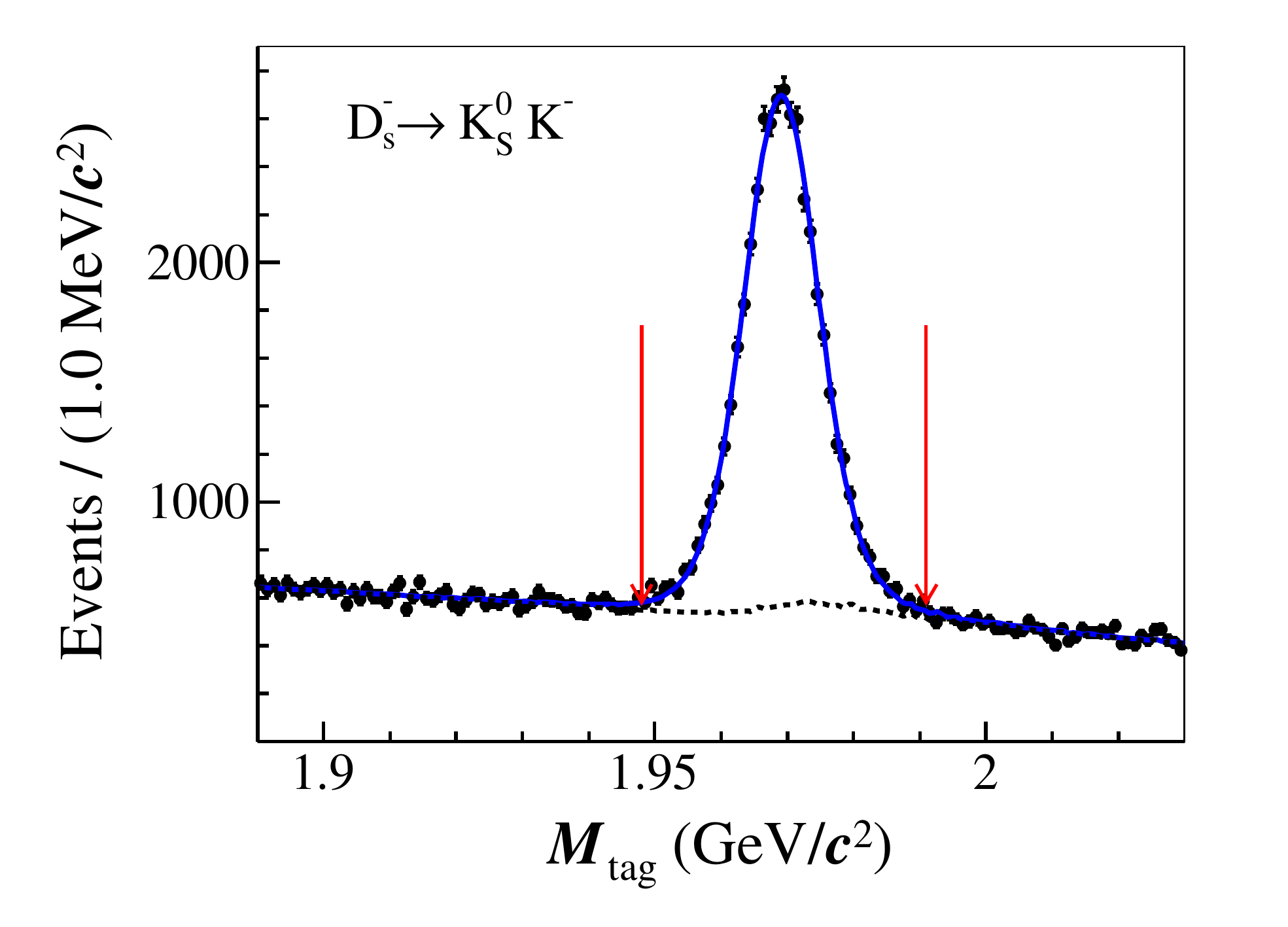}
    \includegraphics[width=6.0cm]{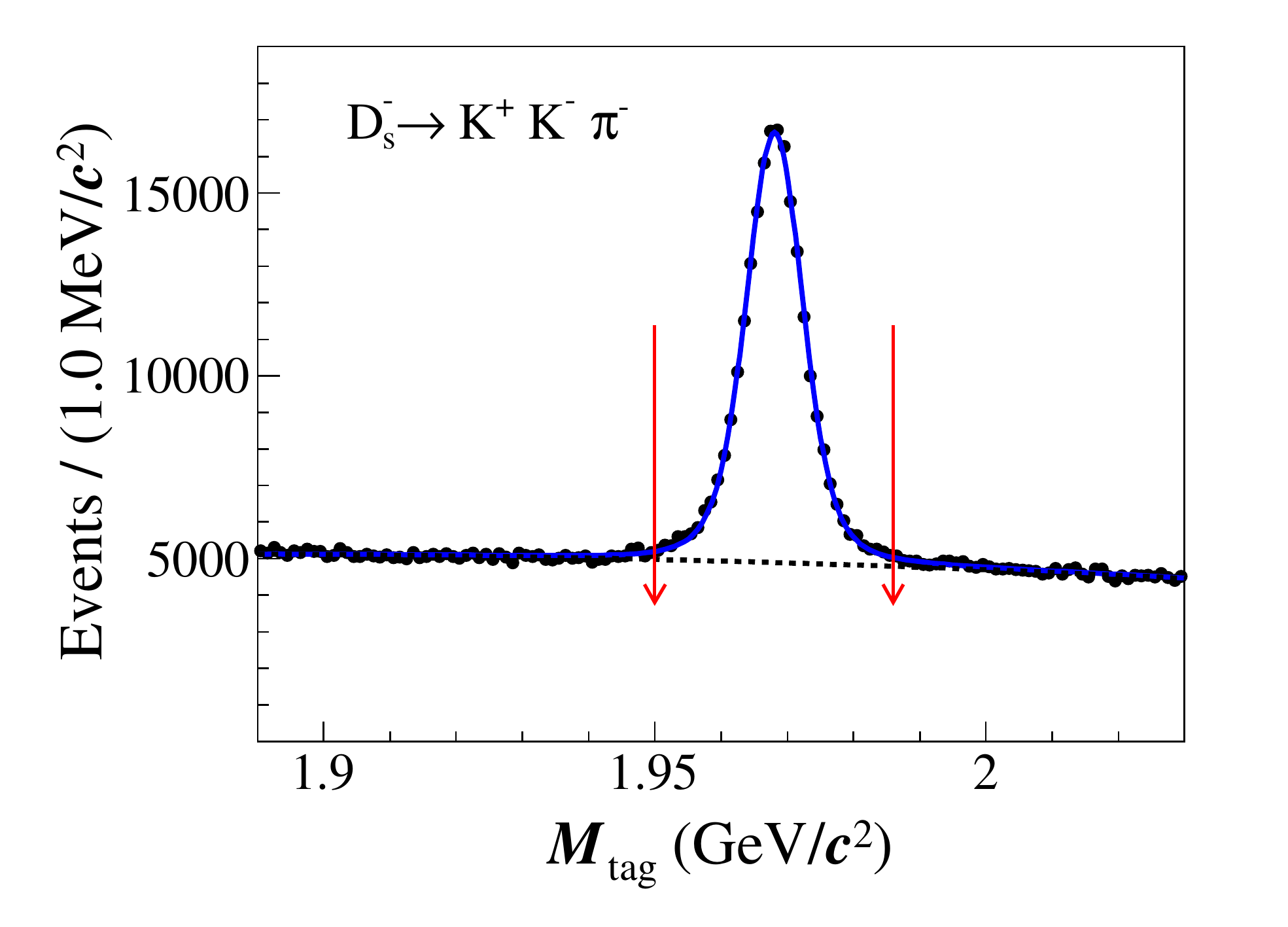}
    \includegraphics[width=6.0cm]{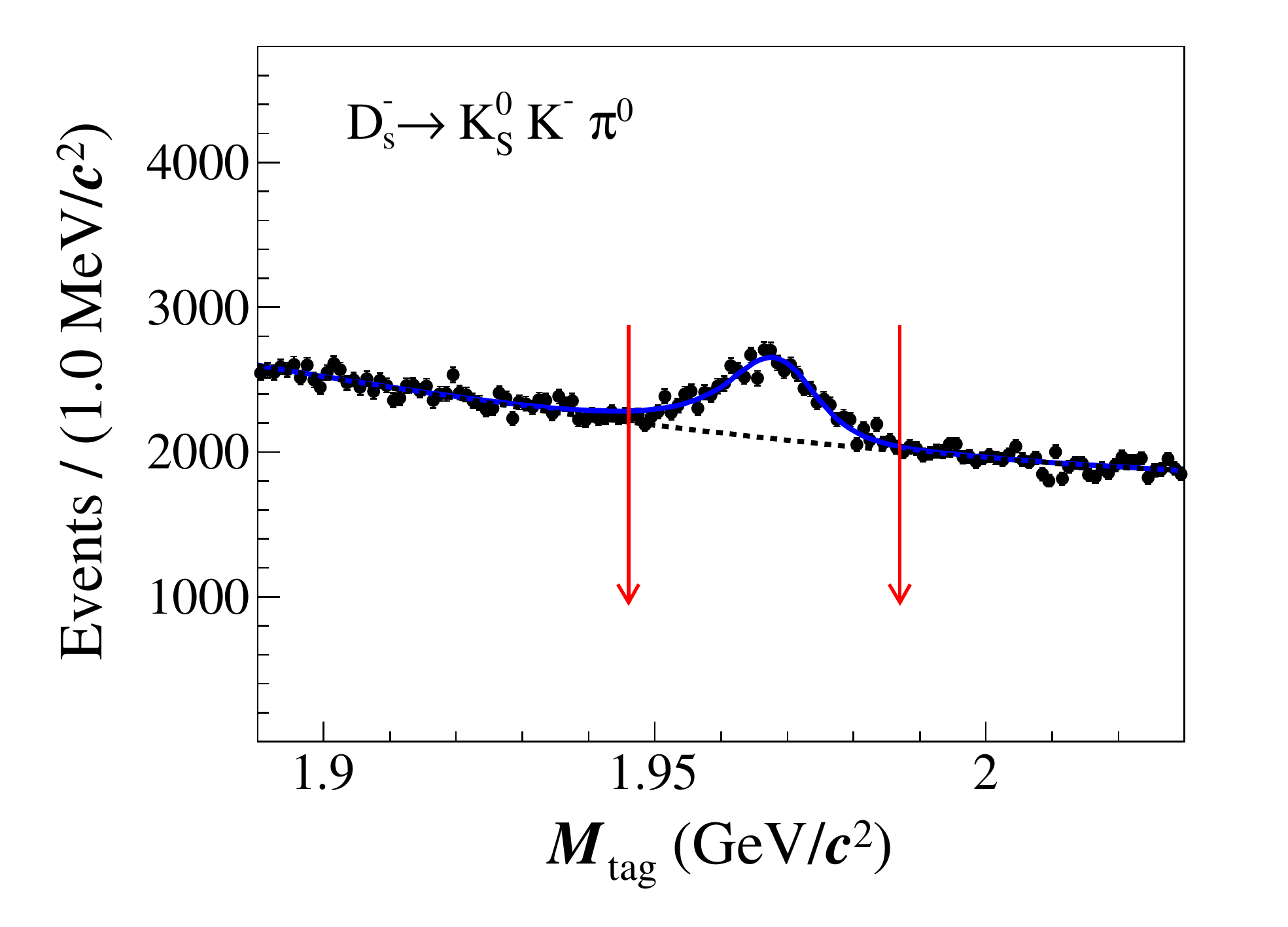}
    \includegraphics[width=6.0cm]{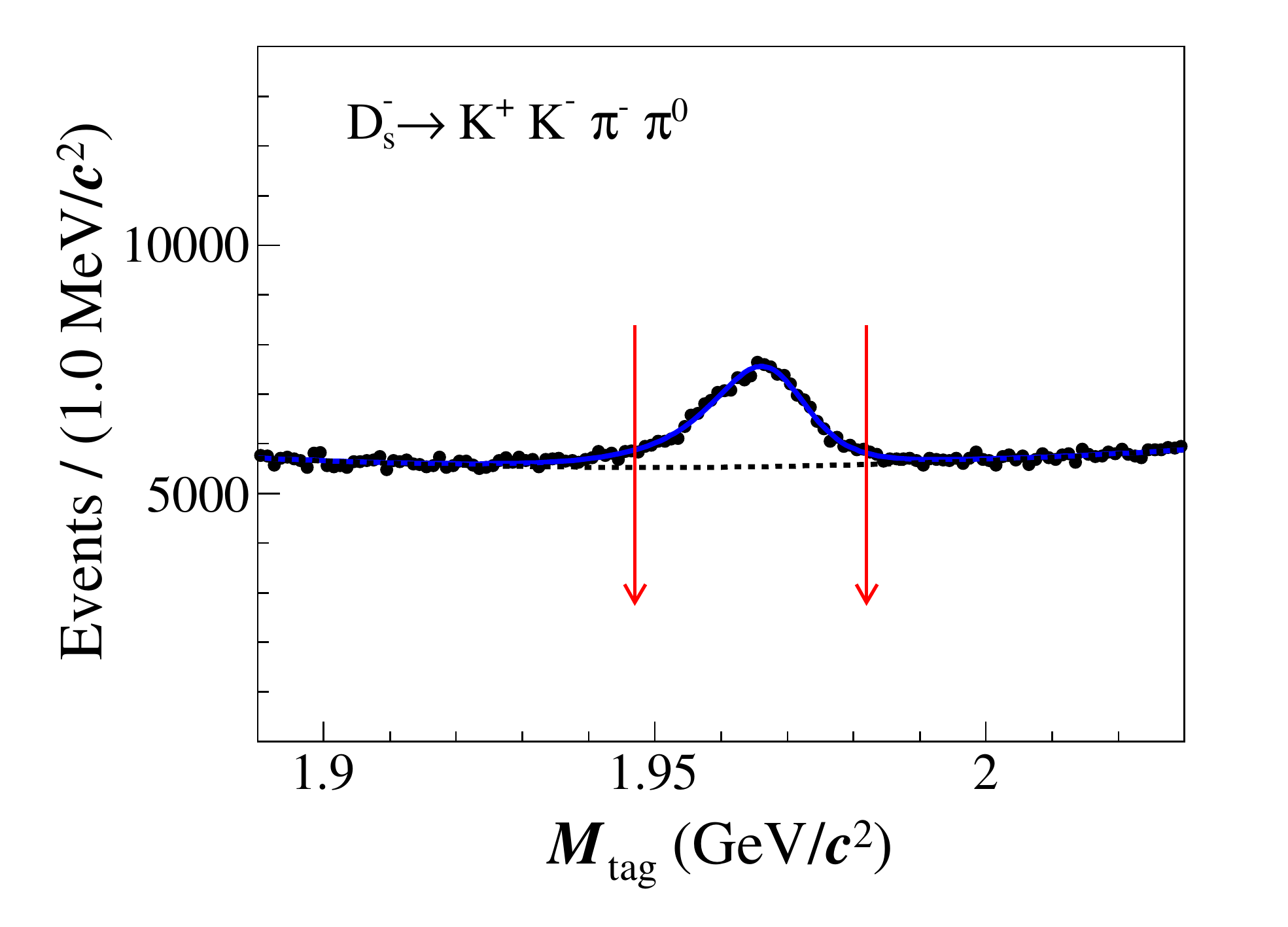}
    \includegraphics[width=6.0cm]{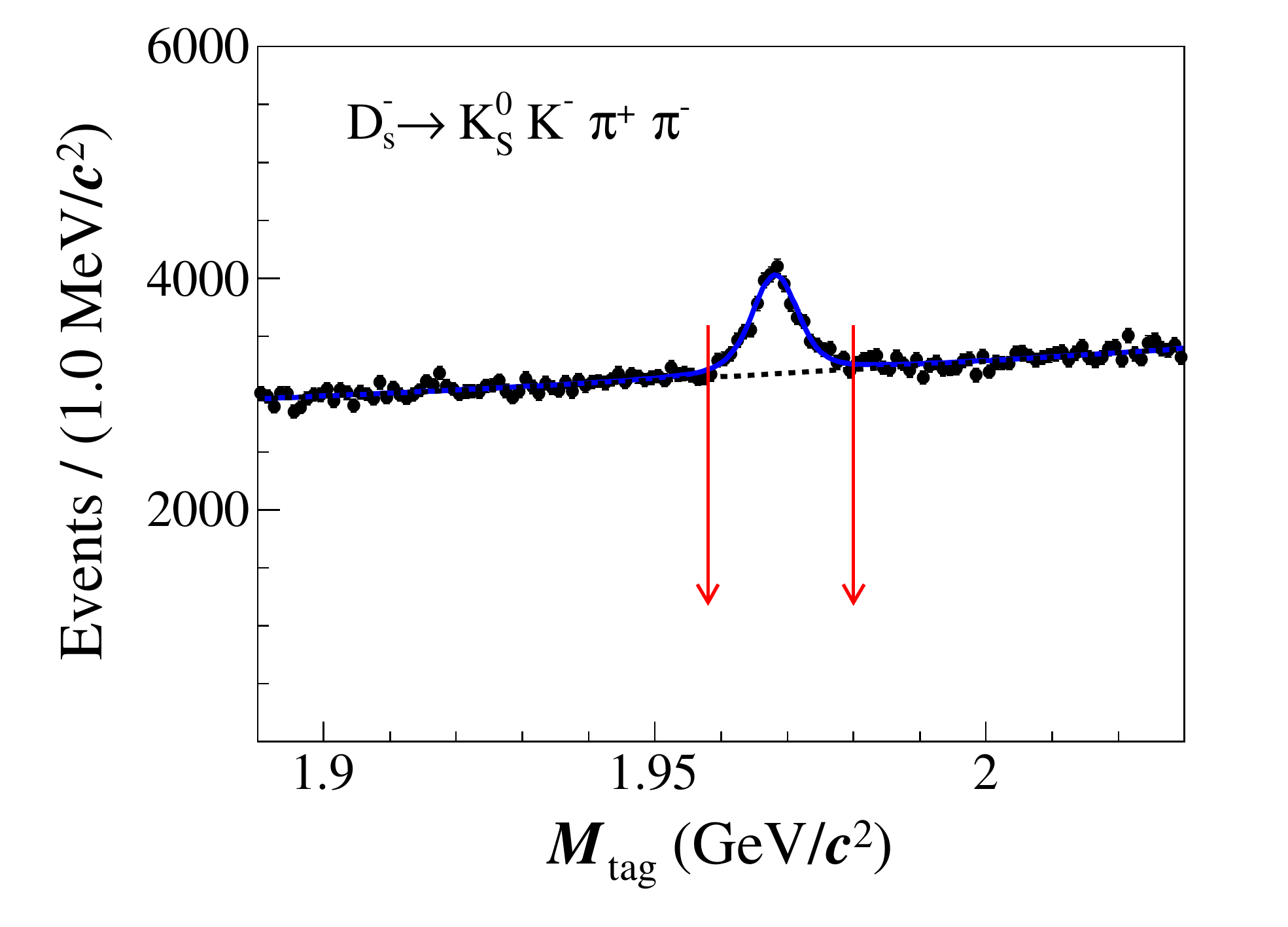}
    \includegraphics[width=6.0cm]{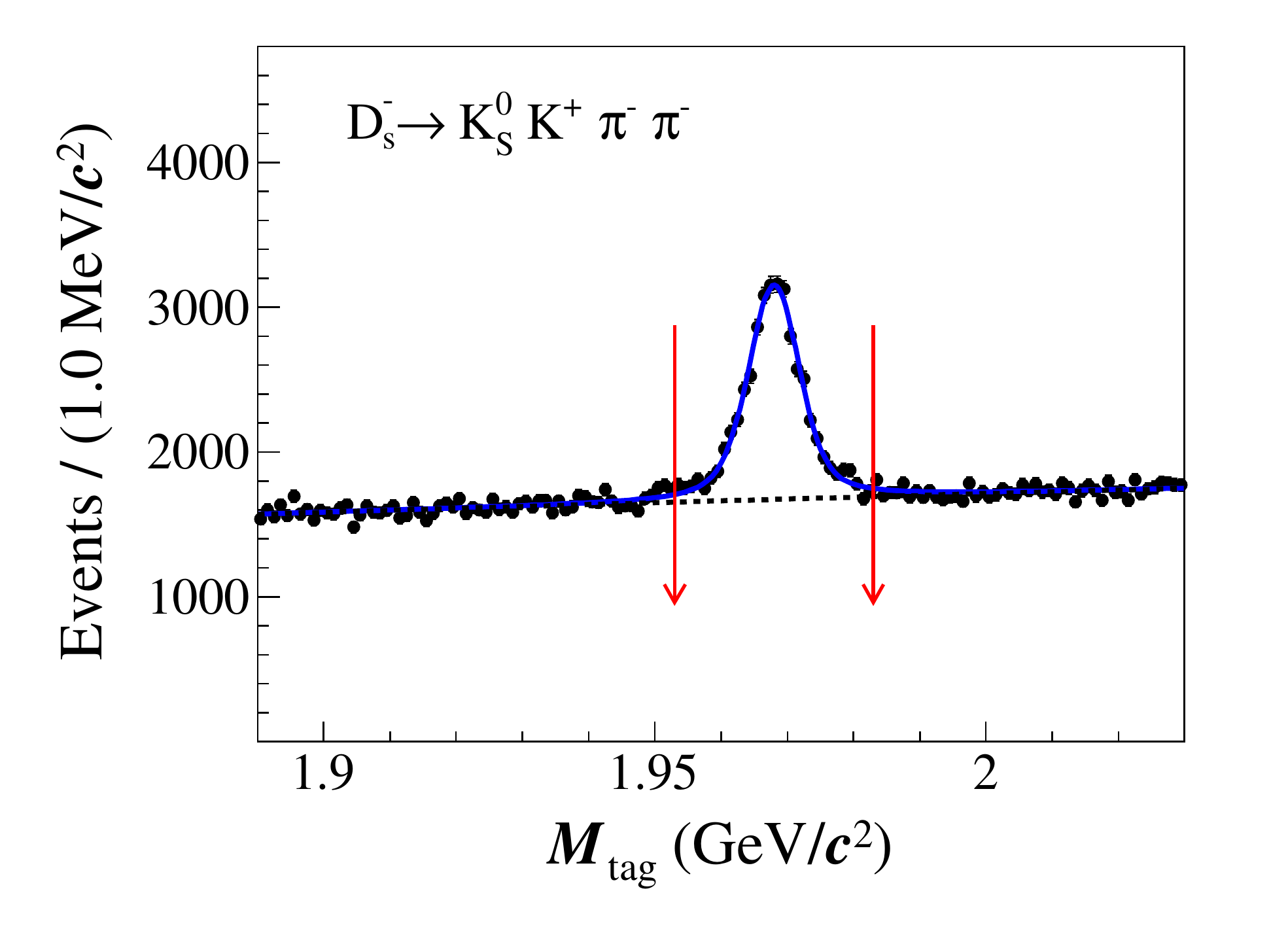}
    \includegraphics[width=6.0cm]{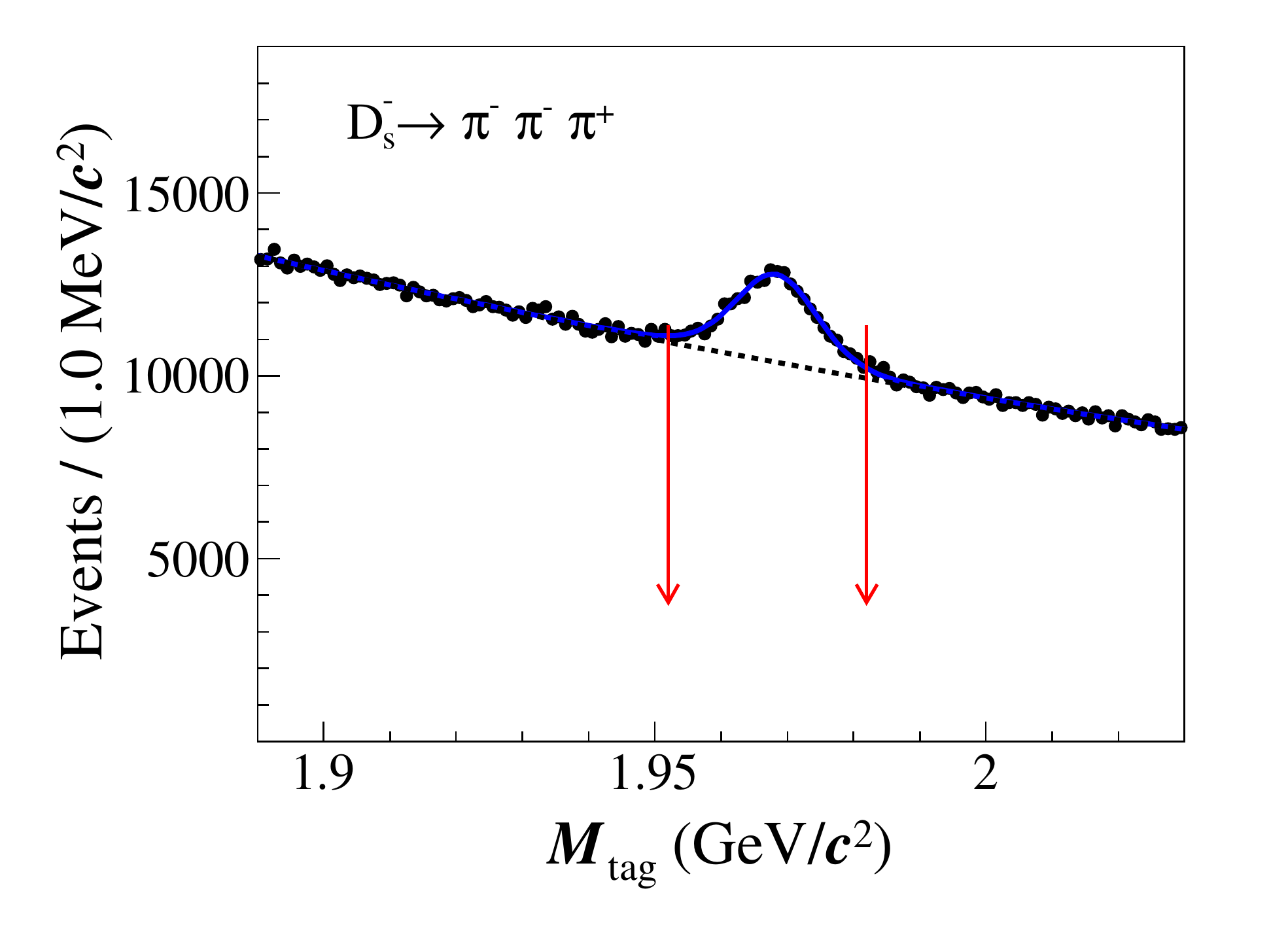}
    \includegraphics[width=6.0cm]{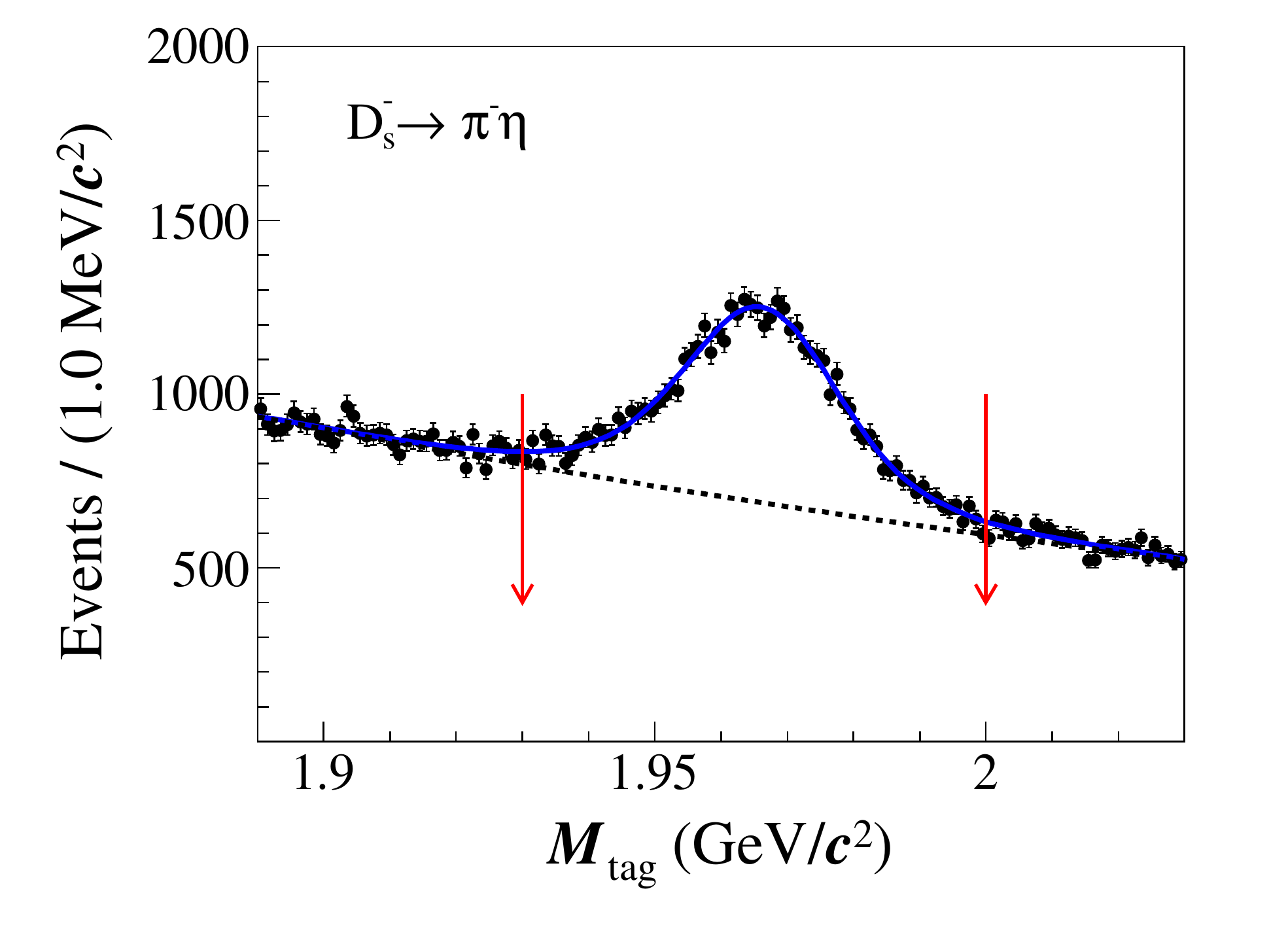}
    \includegraphics[width=6.0cm]{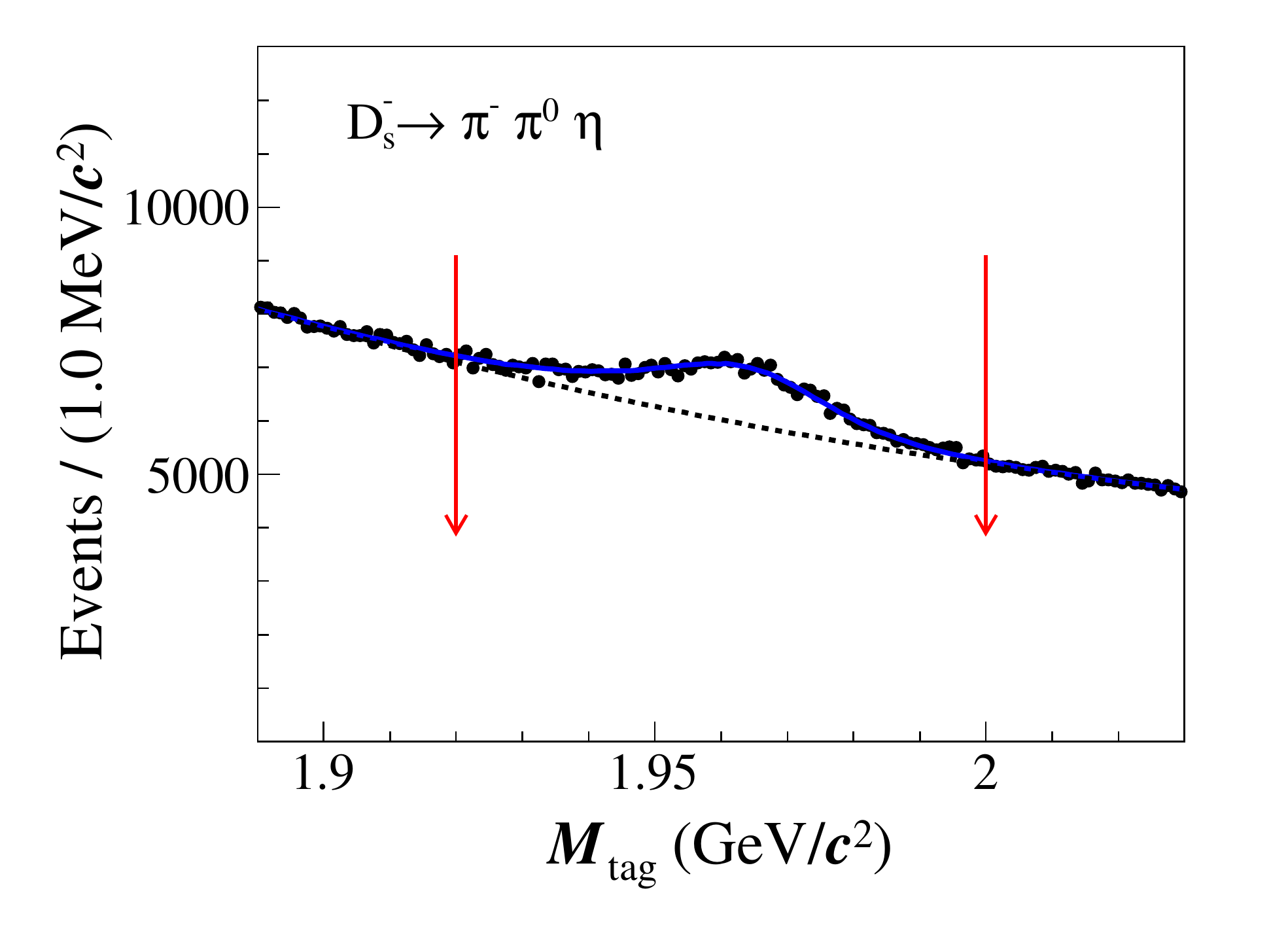}
    \includegraphics[width=6.0cm]{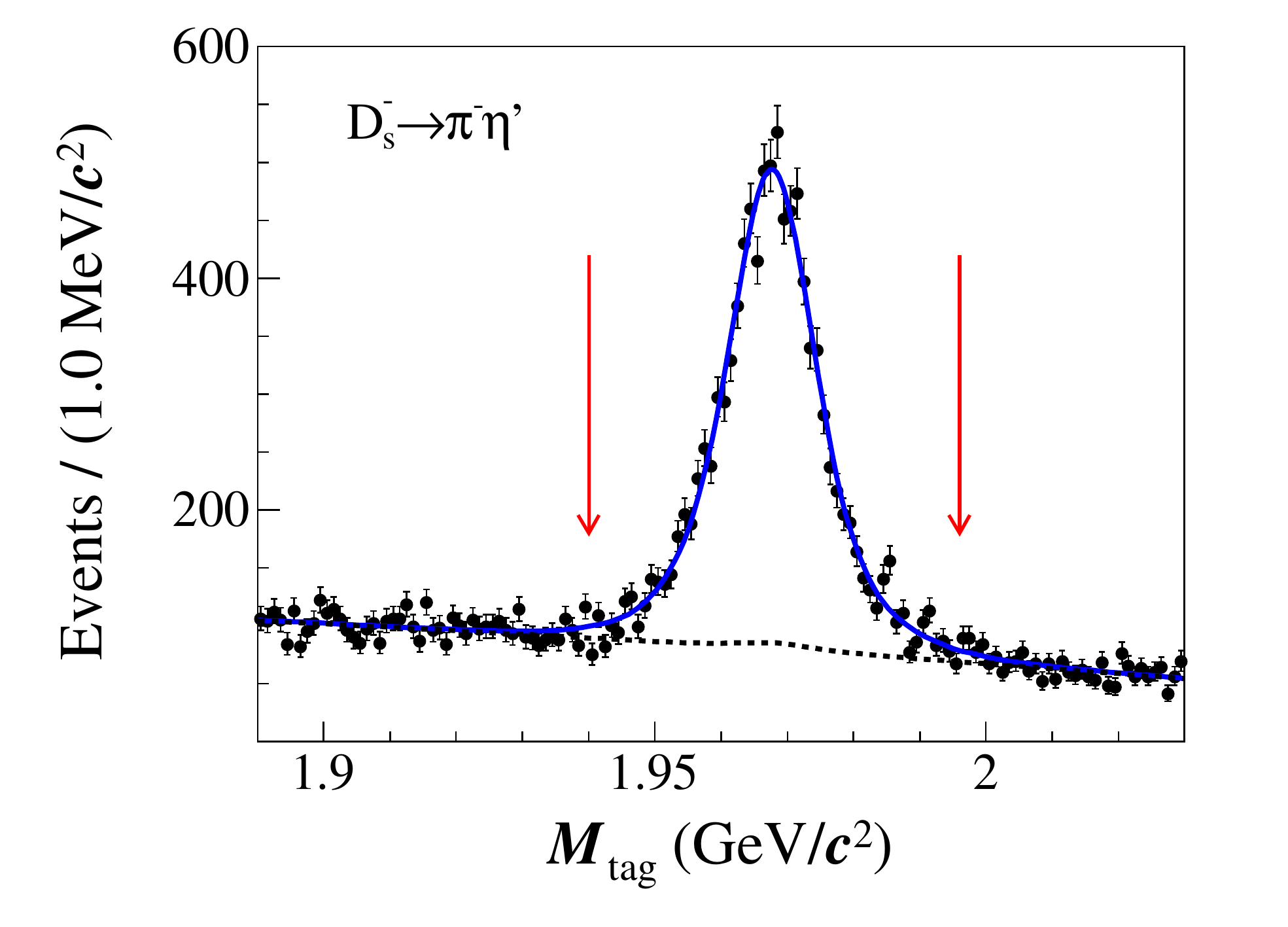}
\caption{Fits to the $M_{\rm tag}$ distributions of the ST candidates from the data sample at $\sqrt{s}=4.178$~GeV. 
	The points with error bars are data, the blue solid lines are the total fits, and the black dashed lines are the fitted backgrounds. 
	The pairs of red arrows denote the signal regions.}
\label{fit:Mass-data-Ds_4180}
\end{center}
\end{figure*}

In order to ensure that the DT sample is a subset of the ST sample in the BF measurement, the ST candidates are selected ahead of the selection of DT candidates. In addition to the selection criteria for final-state particles described in Sec.~\ref{ST-selection}, the requirement $p(\pi) > 100$ MeV$/c$ is applied to all pions in order to exclude transition pions from $D^{*}$ decays.
If there are multiple ST candidates, the combination with the $M_{\rm rec}$ closest to the known mass of $D_s^{*\pm}$~\cite{PDG} is kept. The yields for various tag modes are obtained by fitting the corresponding $M_{\rm tag}$ distributions and listed in Table~\ref{ST-eff}. As an example, the fits to the $M_{\rm tag}$ distributions of the selected ST candidates from the data sample at $\sqrt s=4.178$~GeV are shown in Fig.~\ref{fit:Mass-data-Ds_4180}.
In the fits, the signal is modeled by an MC-simulated shape convolved with a Gaussian function to take into account the data-MC resolution difference. The background is described by a second-order Chebyshev polynomial. 
For the tag modes $D_{s}^{-} \to K_{S}^{0} K^-$ and $D_{s}^{-} \to \pi^-\eta^{\prime}$, there are peaking background contributions coming from $D^{-} \to K_{S}^{0} \pi^-$ and $D_{s}^{-} \to \eta\pi^+\pi^-\pi^-$ decays, respectively. The $D^{-} \to K_{S}^{0} \pi^-$ and $D_{s}^{-} \to \eta\pi^+\pi^-\pi^-$ background are estimated to be $1724\pm34$ and $89\pm5$ events according to the BFs given by PDG~\cite{PDG} and Ref.~\cite{BESIII:2021aza}, corresponding to about $0.3\%$ and less than $0.1\%$ of the total ST yields, respectively.

\begin{table*}[htbp]

    \begin{center}
      \begin{tabular}{l r@{ $\pm$ }l r@{ $\pm$ }l r@{ $\pm$ }l}
        \hline
        Tag mode                                    & \multicolumn{2}{c}{(I) $N_{\rm ST}$}           & \multicolumn{2}{c}{(II) $N_{\rm ST}$}        & \multicolumn{2}{c}{(III) $N_{\rm ST}$}      \\
        \hline
        $D_{s}^{-}\to K_{S}^{0}K^{-}$               & 31941 &312   & 18559 &261 &6582  &160 \\
        $D_{s}^{-}\to K^{+}K^{-}\pi^{-}$            &137240 & 614  & 81286 &505          & 28439 & 327         \\
        $D_{s}^{-}\to K_{S}^{0}K^{-}\pi^{0}$        & 11385&529   & 6832&457  & 2227&220 \\
		$D^{-}_{s}\to K^{+}K^{-}\pi^{-}\pi^{0}$     & 39306&799   & 23311&659            & 7785&453 \\
        $D_{s}^{-}\to K_{S}^{0}K^{-}\pi^{-}\pi^{+}$ & 8093&326   & 5269&282  & 1662&217 \\
        $D_{s}^{-}\to K_{S}^{0}K^{+}\pi^{-}\pi^{-}$ & 15719&289   & 8948&231  & 3263&172 \\
    	$D^{-}_{s}\to \pi^{-}\pi^{-}\pi^{+}$        & 37977&859   & 21909&776            & 7511&393 \\
		$D^{-}_{s}\to \pi^{-}\eta$   & 17940&402   & 10025&339  & 3725&252 \\
		$D_{s}^{-}\to \pi^-\pi^0\eta$          & 42618&1397   & 26067&1196            & 10513&1920 \\
        $D_{s}^{-}\to \pi^{-}\eta^{\prime}$         & 7759&141   & 4428&111  & 1648&74 \\
       
        \hline
      \end{tabular}
    \end{center}
      \caption{The ST yields for the samples collected at $\sqrt{s} =$ (I) 4.178~GeV, (II) 4.189-4.219~GeV,
    and (III) 4.226~GeV. The uncertainties are statistical.}
    \label{ST-eff}
\end{table*}

Once a tag mode is identified, we attempt to reconstruct the signal decay $D_{s}^{+} \to K^{+}\pi^{+}\pi^{-}$. 
If there are multiple candidates, the DT candidate with the average mass, $(M_{\rm sig}+M_{\rm tag})/2$, closest to the $D_{s}^{\pm}$ known mass is retained. A 6C kinematic fit is also performed for the BF measurement, and the same $K_S^0$ veto and $\chi_{\rm 6C}^2$ requirements as in Sec.~\ref{AASelection} are applied to suppress the background.

\begin{figure}[!htbp]
  \centering
  \includegraphics[width=7.cm]{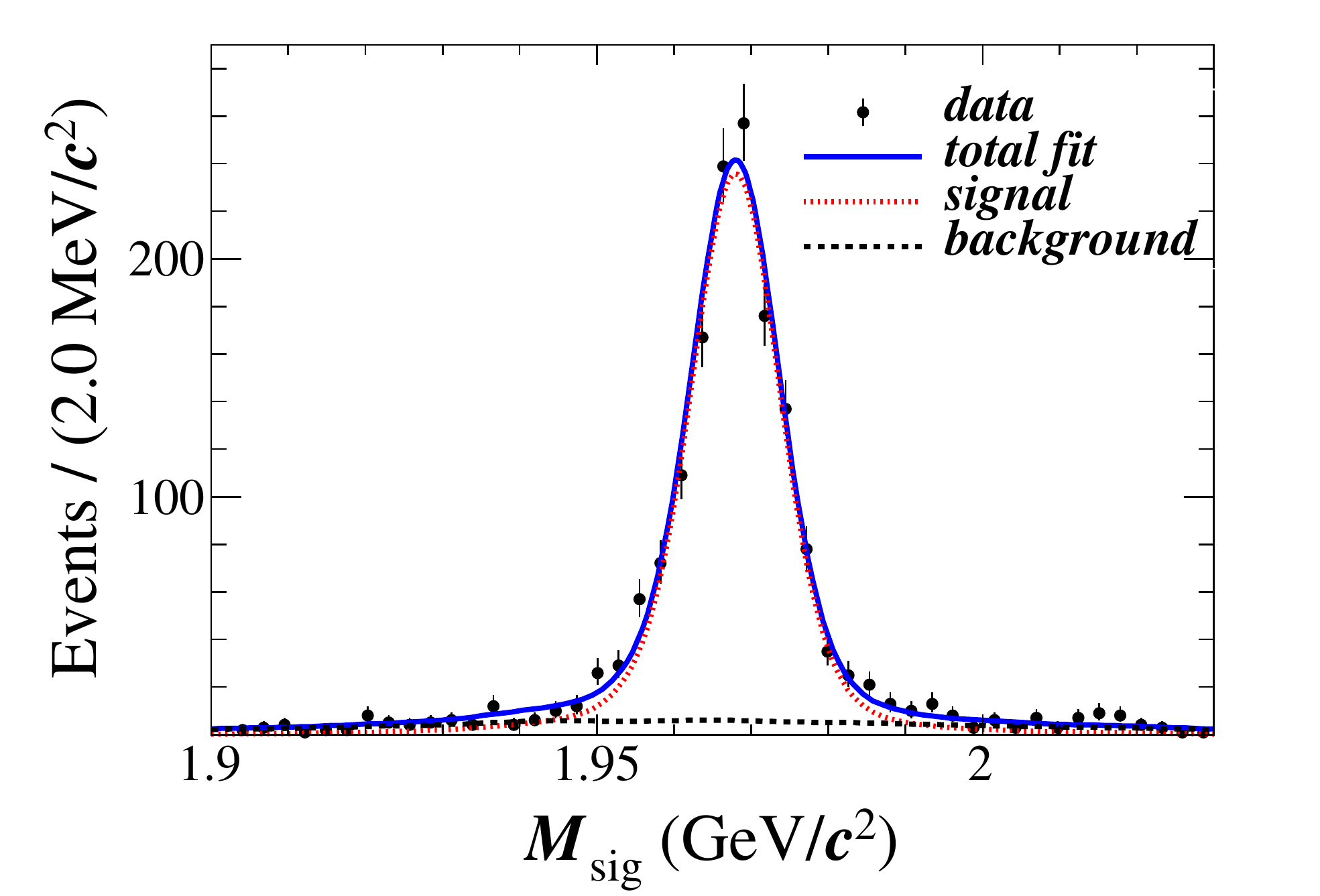}
 \caption{Fit to the $M_{\rm sig}$ distribution of the DT candidates from the data samples at $\sqrt{s}= 4.178$-$4.226$~GeV. 
	The data are represented by points with error bars, the total fit by the blue solid line, and the fitted signal and background by the red dotted and black dashed lines, respectively.}
  \label{DT-fit}
\end{figure}

The DT yield is determined from the fit to the $M_{\rm sig}$ distribution. The fit result is shown in Fig.~\ref{DT-fit}, the signal shape is modeled by an MC-simulated shape convolved with Gaussian function, while the background shape is described with the shape derived from the inclusive MC sample. 
The DT yield obtained is $1415\pm42$. 
Based on this, we determine the BF to be $\mathcal{B}(D^+_s\to K^+\pi^+\pi^-)=(6.11\pm0.18_{\rm stat.}\pm0.11_{\rm syst.})\times 10^{-3}$ taking into account the differences in $K^+$ and $\pi^\pm$ tracking and PID efficiencies between data and MC simulation.

The BFs for the charge-conjugated modes are measured separately. The BFs of $D_s^+ \to K^+\pi^+\pi^-$ and $D_s^- \to K^-\pi^-\pi^+$, denoted as ${\mathcal B}(D^+_s)$ and ${\mathcal B}(D^-_s)$, are measured to be $(5.88\pm0.25_{\rm stat.}\pm0.11_{\rm syst.})\times 10^{-3}$ and $(6.28\pm0.26_{\rm stat.}\pm0.11_{\rm syst.})\times 10^{-3}$, respectively. The asymmetry of the two BFs is determined to be $A_{\mathit{{CP}}} = \frac{\mathcal{B}(D_s^+)-\mathcal{B}(D_s^-)}{\mathcal{B}(D_s^+)+\mathcal{B}(D_s^-)} = (3.3\pm{{3.0}}_{\rm stat.}\pm1.3_{\rm syst.})\%$. The systematic uncertainties of tracking and PID have been canceled in the $A_{\mathit{{CP}}}$ calculation. The result is consistent with the hypothesis of $\mathit{{CP}}$ symmetry~\cite{PhysRevD.89.054006}.

The systematic uncertainties in the BF measurement are discussed below.
\begin{itemize}
    \item ST yield. The uncertainty of the total yield of the ST $D_s^-$ mesons is determined to be 0.5$\%$ by taking into account the background fluctuation in the fit, and examining the changes of the fit yields when varying the background shape.
    \item Background shape. To estimate the uncertainty due to the background shape of the signal $D_s^+$ invariant mass distribution, a second-order Chebychev polynomial is used to replace the MC-simulated shape, and an uncertainty of 0.6\% is obtained.
    \item Tracking and PID. The processes $e^+e^- \to K^+K^-K^+K^-$, $K^+K^-\pi^+\pi^-(\pi^0)$ and $ \pi^+\pi^-\pi^+\pi^-(\pi^0)$ are used to study the tracking and PID efficiencies of $K^+$ and $\pi^\pm$. 
    The data-MC tracking and PID efficiencies ratios of $\pi^+(\pi^-)$ are $0.998\pm0.003\ (1.002\pm0.003)$ and $1.002\pm0.002\ (1.003\pm0.002)$, respectively. The data-MC tracking and PID efficiencies ratios of $K^+ (K^-)$ are $0.998\pm0.003\ (0.997\pm0.003)$ and $1.003\pm0.002\ (1.003\pm0.002)$, respectively. Finally,
    the systematic uncertainties associated with tracking and PID efficiencies for each charged particle are estimated to be 0.3\% and 0.2\%, respectively.
    \item MC sample size. The uncertainty of the MC sample size is given by $\sqrt{\sum_\alpha{(\frac{f_\alpha\delta_{\epsilon_\alpha}}{\epsilon_\alpha})^2}}$, where $f_\alpha$ is the tag yield fraction and $\epsilon_\alpha$ is the average DT efficiency of tag mode $\alpha$. The corresponding uncertainty is determined to be 0.4\%.
    \item Amplitude model. 
    The uncertainty from the amplitude model is determined by varying the amplitude model parameters based on their error matrix 600 times. A Gaussian function is used to fit the distribution of 600 DT efficiencies and the fitted width divided by the mean value is taken as an uncertainty. The related uncertainty is 0.5\%.
    
\item $\chi^2_{\rm 6C}$ requirement. The uncertainty of the $\chi^2_{\rm 6C}$ requirement is assigned to be the difference between the data and MC efficiencies of the $D_s^+ \to K^+K^-\pi^+$ candidates. The data and MC simulation control samples of $D_s^+ \to K^+K^-\pi^+$ including over 99$\%$ signal events are selected. Then, the efficiency corresponding to the $\chi^2_{\rm 6C}$ requirement is obtained, and the uncertainty is calculated by $1-\frac{\varepsilon_{\rm data}}{\varepsilon_{\rm MC}}$, where $\varepsilon_{\rm data}$ and $\varepsilon_{\rm MC}$ are the selection efficiencies of data and MC simulation, respectively. The associated systematic uncertainty is assigned to be 1.0\%. 
\item $K_S^0$ rejection. The uncertainty of $K_S^0$ rejection has been included in the uncertainty of the amplitude model, in which the inconsistency of the structure of $\pi^+ \pi^-$ spectrum between data and MC is estimated by varying the amplitude model parameters by 600 times. On the other hand, the remained $D_s^+ \to K_S^0K^+$ events are less than 0.1\% after $K_S^0$ rejection, thus the systematic uncertainty of $K_S^0$ rejection can be negligible. 

\end{itemize}

All of the systematic uncertainties are summarized in Table~\ref{BF-Sys}.
Adding them in quadrature results in a total systematic uncertainty of 1.8\% in the BF measurement.

\begin{table}[htbp]

  \centering
  \begin{tabular}{lc}
\hline
Source            & Uncertainty (\%) \\
\hline
ST yield          & 0.5\\
Background shape  & 0.6\\
Tracking          & 0.9\\
PID               & 0.6\\
MC sample size    & 0.4\\
Amplitude model   & 0.5\\
$\chi^2_{\rm 6C}$ requirement & 1.0\\
\hline
Total             & 1.8\\
\hline
  \end{tabular}
    \caption{Systematic uncertainties in the BF measurement.}
  \label{BF-Sys}
\end{table}

\section{Summary}
Using $e^+e^-$ collision data equivalent to an integrated luminosity of $\rm 6.32\ fb^{-1}$ recorded with the BESIII detector at the center-of-mass energies between 4.178 and 4.226 GeV, an amplitude analysis of the decay $D^+_s\to K^+\pi^+\pi^-$ has been performed. 
The results for the FFs and phases of the different intermediate processes are listed in Table~\ref{tab:signi}. 
The BF for the decay $D^+_s\to K^+\pi^+\pi^-$ is measured to be $(6.11\pm0.18_{\rm stat.}\pm0.11_{\rm syst.})\times 10^{-3}$, which is improved by about a factor of 2 compared to the world average value~\cite{PDG}. 
The BFs for the intermediate processes calculated with $\mathcal{B}_i = {\rm FF}_i \times \mathcal{B}(D^+_s\to K^+\pi^+\pi^-)$ in this analysis and from the PDG~\cite{PDG} are listed in Table~\ref{inter-processes}. The BFs of $D_s^+ \to K^+f_0(500)$, $D_s^+ \to K^+f_0(980)$, and $D_s^+ \to K^+f_0(1370)$ are determined for the first time. The asymmetry of the BFs of $D_s^+ \to K^+\pi^+\pi^-$ and $D_s^- \to K^-\pi^-\pi^+$ is determined to be $(3.3\pm{{3.0}}_{\rm stat.}\pm1.3_{\rm syst.})\%$. No indication of $\mathit{{CP}}$ violation is found. 

The obtained BF of $D_s^+ \to K^+\rho^0$ is in good agreement with the predictions in Ref.~\cite{PhysRevD.89.054006}, and the measured BF of $D_s^+ \to K^*(892)^0\pi^+$ is consistent with the prediction in Ref.~\cite{Cheng:2016ejf}. Meanwhile, our result deviates from the predictions of $D_s^+ \to K^+\rho^0$ in Refs.~\cite{Cheng:2016ejf,Wu:2004ht} and $D_s^+ \to K^*(892)^0\pi^+$ in Refs.~\cite{Wu:2004ht,PhysRevD.89.054006} over two standard deviations. {{Moreover,
Ref.~\cite{PhysRevD.89.054006} predicts the ratio of BF of $D_s^+ \to K^+\rho^0$ to that of $D_s^+ \to K^+\omega$ is far greater than one, while Ref.~\cite{Cheng:2016ejf} calculates that it should be close to one. The ratio is determined to be about two by taking the results in this analysis and in Ref.~\cite{BESIII:2022bvv}.}} More precise theoretical predictions are desirable to understand the $D_s^{\pm} \to VP$ processes {{and ${\rm SU(3)}_F$ flavor symmetry breaking effect}}.

\begin{table}[htbp]

	\centering
	\begin{tabular}{l c@{ $\pm$ }c@{ $\pm$ }c c@{ $\pm$ }c}
	\hline
		Intermediate process  &\multicolumn{3}{c}{BF($10^{-3}$)} &\multicolumn{2}{c}{PDG($10^{-3}$)}\\
	\hline
		$D_s^{+}\to K^+\rho^0$  &{{1.96}} &{{0.19}} &{{0.23}} &2.5 &0.4\\
		$D_s^{+}\to K^+\rho(1450)^0$ &{{0.80}} &{{0.19}} & {{0.18}} &0.69 &0.64\\
		$D_s^{+}\to K^*(892)^0\pi^+$ &1.85 &{{0.12}} & {{0.13}} &1.41 &0.24\\
		$D_s^{+}\to K^*(1410)^0\pi^+$ &{{0.27}} & 0.13 & {{0.15}} &1.23 &0.28 \\
		$D_s^{+}\to K^*_0(1430)^0\pi^+$ &{{1.13}} & 0.16 &{{0.16}} &0.50 &0.35\\
		$D_s^{+}\to K^+f_0(500)$ &{{0.44}} &{{0.13}} &{{0.27}} &\multicolumn{2}{c}{-}\\
		$D_s^{+}\to K^+f_0(980)$ &0.27 & 0.08 & {0.07} &\multicolumn{2}{c}{-}\\
		$D_s^{+}\to K^+f_0(1370)$ &1.22 &{{0.18}} & {{0.57}} &\multicolumn{2}{c}{-}\\
		$D_s^{+}\to (K^+\pi^+\pi^-)_{NR}$ &\multicolumn{3}{c}{-} &1.03 &0.34\\
        \hline
	\end{tabular}
	\caption{The BFs for various intermediate processes decaying into the final state $K^+\pi^+\pi^-$ in this analysis and from the PDG~\cite{PDG}, where the first and second uncertainties are statistical and systematic, respectively.}
	\label{inter-processes}
\end{table}

\acknowledgments
The BESIII collaboration thanks the staff of BEPCII and the IHEP computing center for their strong support. This work is supported in part by National Key Research and Development Program of China under Contracts Nos. 2020YFA0406400, 2020YFA0406300; National Natural Science Foundation of China (NSFC) under Contracts Nos. 11625523, 11635010, 11735014, 11775027, 11822506, 11835012, 11875054, 11935015, 11935016, 11935018, 11961141012, 12192260, 12192261, 12192262, 12192263, 12192264, 12192265; the Chinese Academy of Sciences (CAS) Large-Scale Scientific Facility Program; Joint Large-Scale Scientific Facility Funds of the NSFC and CAS under Contracts Nos. U1732263, U1832207, U2032104; CAS Key Research Program of Frontier Sciences under Contracts Nos. QYZDJ-SSW-SLH003, QYZDJ-SSW-SLH040; 100 Talents Program of CAS; INPAC and Shanghai Key Laboratory for Particle Physics and Cosmology; Natural Science Foundation of Hunan Province, China under Grant No. 2021JJ40036; ERC under Contract No. 758462; European Union Horizon 2020 research and innovation programme under Contract No. Marie Sklodowska-Curie grant agreement No 894790; German Research Foundation DFG under Contracts Nos. 443159800, Collaborative Research Center CRC 1044, FOR 2359, FOR 2359, GRK 214; Istituto Nazionale di Fisica Nucleare, Italy; Ministry of Development of Turkey under Contract No. DPT2006K-120470; National Science and Technology fund; Olle Engkvist Foundation under Contract No. 200-0605; STFC (United Kingdom); The Knut and Alice Wallenberg Foundation (Sweden) under Contract No. 2016.0157; The Royal Society, UK under Contracts Nos. DH140054, DH160214; The Swedish Research Council; U. S. Department of Energy under Contracts Nos. DE-FG02-05ER41374, DE-SC-0012069.

\bibliographystyle{JHEP}
\bibliography{references}

\clearpage
\appendix
\large
The BESIII Collaboration\\
\normalsize
\\M.~Ablikim$^{1}$, M.~N.~Achasov$^{10,b}$, P.~Adlarson$^{69}$, M.~Albrecht$^{4}$, R.~Aliberti$^{29}$, A.~Amoroso$^{68A,68C}$, M.~R.~An$^{33}$, Q.~An$^{65,51}$, X.~H.~Bai$^{59}$, Y.~Bai$^{50}$
% ...
\\
\vspace{0.2cm}
\\
\vspace{0.2cm} {\it
$^{1}$ Institute of High Energy Physics, Beijing 100049, People's Republic of China\\
% ...
$^{4}$ Bochum Ruhr-University, D-44780 Bochum, Germany\\
% ...
$^{10}$ G.I. Budker Institute of Nuclear Physics SB RAS (BINP), Novosibirsk 630090, Russia\\
% ...
$^{29}$ Johannes Gutenberg University of Mainz, Johann-Joachim-Becher-Weg 45, D-55099 Mainz, Germany\\
% ...
$^{33}$ Liaoning Normal University, Dalian 116029, People's Republic of China\\
% ...
$^{50}$ Southeast University, Nanjing 211100, People's Republic of China\\
$^{51}$ State Key Laboratory of Particle Detection and Electronics, Beijing 100049, Hefei 230026, People's Republic of China\\
% ...
$^{59}$ University of Jinan, Jinan 250022, People's Republic of China\\
% ...
$^{65}$ University of Science and Technology of China, Hefei 230026, People's Republic of China\\
% ...
$^{68}$ University of Turin and INFN, (A)University of Turin, I-10125, Turin, Italy; (B)University of Eastern Piedmont, I-15121, Alessandria, Italy; (C)INFN, I-10125, Turin, Italy\\
$^{69}$ Uppsala University, Box 516, SE-75120 Uppsala, Sweden\\
% ...
$^{b}$ Also at the Novosibirsk State University, Novosibirsk, 630090, Russia\\
% ...
}


\end{document}